\documentclass[aps,preprint,amsfonts]{revtex4}
\usepackage[]{bm,epsfig}
\def\br{\bbox{r}}
\def\bL{\bbox{L}}

\newcommand{\be}{\begin{equation}}
\newcommand{\ee}{\end{equation}}

\newcommand{\JCP}[1]{J. Chem. Phys.\ {\bf #1}}

\newcommand{\PRA}[1]{Phys. Rev. A\ {\bf #1}}

\begin{document}

\title{
Sympathetic cooling of the Ba$^+$ ion by collisions
with ultracold Rb atoms: theoretical prospects}

\author{\sc Micha\l\ Krych}

\affiliation{\sl Institute of Theoretical Physics, Department of Physics, University of Warsaw, Ho\.za 69,
00-681 Warsaw, Poland and Quantum Chemistry Laboratory, Department of Chemistry, University of Warsaw, Pasteura 1,
02-093 Warsaw, Poland}

\author{\sc Wojciech Skomorowski, Filip Paw\l owski, 
Robert Moszynski\footnote[1]{Author for correspondence;
e-mail:robert.moszynski@tiger.chem.uw.edu.pl}}

\affiliation{\sl Quantum Chemistry Laboratory, Department of Chemistry, University of Warsaw, Pasteura 1,
02-093 Warsaw, Poland}

\author{\sc Zbigniew Idziaszek}

\affiliation{\sl Institute of Theoretical Physics,  Department of Physics, University of Warsaw, Ho\.za 69,
00-681 Warsaw, Poland and Institut de Physique de Rennes, UMR 6251 du CNRS and Universit\'e de Rennes I,
35042 Rennes Cedex, France}

\begin{abstract}
State-of-the-art {\em ab initio} techniques have been applied to compute the
potential energy curves of the (BaRb)$^+$ molecular ion in the Born-Oppenheimer
approximation
for the singlet and triplet states dissociating into the ground state $^1$S Rb$^+$
ion
and the Ba atom in the ground $^1$S state, the lowest singlet or triplet D excited states, and
for the singlet and triplet states dissociating into the ground state $^2$S Rb atom
and the ground state $^2$S Ba$^+$ ion. The ground state potential energy was
obtained with the coupled cluster method restricted to single, double, and
nonperturbative triple excitations. The first triplet states in the $\Sigma$,
$\Pi$, and $\Delta$ symmetries were computed with the restricted open-shell
coupled cluster method restricted to single, double, and
nonperturbative triple excitations.
All other excited state potential energy curves were computed using
the equation of motion approach
within the coupled-cluster singles, doubles, and linear triples framework.
The long-range coefficients describing the electrostatic, induction, and
dispersion interactions at large interatomic distances are also reported.
The electric transition dipole moments
governing the X$^1\Sigma\; \rightarrow\; ^1\Sigma, \; ^1\Pi$
have been obtained as the first residue of the
polarization propagator computed with the linear response coupled-cluster
method restricted to single and double excitations.
Nonadiabatic radial and angular coupling matrix elements, as well as the
spin-orbit coupling matrix elements have been evaluated using the
multireference configuration interaction method restricted to single
and double excitations with a large active space.
With these couplings, the spin-orbit coupled (relativistic) potential energy
curves for the 0$^+$ and 1 states relevant for the running experiments
have been obtained. Finally, relativistic transition moments and
nonadiabatic coupling matrix elements were obtained from the
nonrelativistic results and spin-orbit eigenvectors.
The electronic structure input has been employed in the single channel
scattering calculations of the collisional cross sections between
the Ba$^+$ ion and Rb atom. Both nonrelativistic and relativistic
potentials were used in these calculations. Our results show that
the inelastic cross section corresponding to the charge transfer from
the Rb atom to the Ba$^+$ ion is much smaller than the elastic one over
a wide range of energies up to 1 mK. This suggests that sympathetic
cooling of the Ba$^+$ ion by collisions with ultracold Rb atoms should
be possible.
\end{abstract}

\pacs{34.20.Cf, 34.70.+e, 34.50.Cx}

\maketitle
\newpage

\section{Introduction}
\label{sec1}

Nowadays an increasing number of experimental groups worldwide start to work with hybrid systems involving cold or ultracold trapped atoms and ions \cite{Vuletic2009,Smith2005,Zipkes2010,Zipkes2010a,Schmid}. Apart from the fundamental interest in the physics of atom-ion collisions in the quantum regime \cite{Cote2000,Bodo2008,Idziaszek2009}, these new systems are very attractive from the point of view of quantum information processing \cite{Idziaszek2007,Doerk2010}, studying many-body effects of ion impurities \cite{Massignan2005}, or creation of the molecular ions \cite{Cote2003}. One type of ongoing experiments is focused on studying the collisional processes in cold clouds of atoms and ions, like Yb with Yb$^{+}$ and Na with Ca$^{+}$ \cite{Vuletic2009,Smith2005} stored in dual hybrid charged-neutral traps at mK temperatures. Other experiments, in contrast, study the dynamics of single ions like Ba$^+$ or Yb$^+$ trapped in RF potentials and immersed in Bose-Einstein condensates \cite{Zipkes2010,Zipkes2010a,Schmid}. In such experiments the ultracold cloud of atoms is prepared using standard cooling and trapping techniques, while the ion is laser cooled separately in RF trap and later overlapped with the Bose-condensed atoms. The atom-ion collisions can lead to further sympathetic cooling of the ion, but the net cooling effect depends on the interplay of $(i)$  the two-body collisional properties, $(ii)$ the micromotion of the ion due to the time-dependent RF potential, $(iii)$ the collective phenomena resulting from the coherent properties of the condensate additionally modified by the presence of an ionic impurity. In this paper we perform a first step towards understanding the sympathetic cooling process of Ba$^+$ ion with Rb atoms, by calculating highly accurate molecular potentials and determing the single channel elastic, spin-exchange and charge-transfer collision rates.

In a recent paper Makarov et al. \cite{Cote:03} performed {\em ab initio} and
dynamical calculations on the (CaNa)$^+$ molecular ion. The results of these
calculations suggest that the milliKelvin regime of collisional cooling
of calcium ions by sodium atoms is very favorable, with the rate coefficient
for charge transfer from the Na atom to the Ca$^+$ ion several orders of
magnitude smaller than the rate for elastic and spin-exchange collisions.
This strongly suggests that sympathetic cooling of ions by collisions with
ultracold atoms should be possible. This system was further studied in the
ultracold regime in Ref. \cite{Idziaszek2009}
within the multichannel quantum defect formalism. In the present paper we
investigate the possibility of sympathetic cooling for yet another system of experimental interest \cite{Schmid}: Ba$^+$  ions cooled by collisions with ultracold Rb atoms.

%However, as mentioned above the running and forthcoming experiments will mostly concern Yb$^+$ \cite{yb+exp1,yb+exp2} and Ba$^+$ \cite{ba+exp} ions cooled by collisions with ultracold Rb atoms.

Theoretical modeling of collisions in the ultracold regime requires a lot of care
\cite{buss0,buss1,buss2,buss3,Koch:08}. First of all, the electronic states involved
in the dynamics must be computed with the state-of-the-art methods of quantum chemistry.
In particular, these methods should be size-consistent in order to ensure a proper
dissociation of the molecular state, and must sufficiently account for the electronic
correlation. Moreover, any information on the long-range
asymptotics of the potentials is very important. Finally, all couplings
between the molecular states, both those
resulting from the spin-orbit interaction and from the nonadiabatic effects,
should be considered. Having the electronic structure results at hand, exact
quantum-dynamical calculations should be performed to get the cross sections
and collisional rates.

In a recent series of papers Knecht et al. \cite{Knecht:10,Knecht:10a} reported
nonrelativistic, scalar relativistic, and fully relativistic {\em ab initio}
potential energy curves for the (BaRb)$^+$ molecular ion. Unfortunately, the
approach adopted in these papers is not size-consistent, so the results at large
internuclear distances may not be accurate enough. Moreover, excitations in the
wave function beyond the doubles levels were included only for the simplest
single reference closed-shell and high-spin states. Also, the transition
moments between states, necessary to model the radiative charge transfer process
from the Rb atom to the Ba$^+$ ion, were not computed. Finally, the nonadiabatic
coupling matrix elements between different molecular states were not considered.
We believe that such a level of {\em ab initio} calculations is not sufficient
to properly model ultracold collisions in the (BaRb)$^+$ molecular ion.
Therefore, in the present paper we report state-of-the-art {\em ab initio}
potential energy curves for the (BaRb)$^+$ molecular ion in the Born-Oppenheimer
approximation
for the singlet and triplet states dissociating into the ground state $^1$S Rb$^+$
ion
and the Ba atom in the ground $^1$S state,
the lowest singlet or triplet D excited states, and
for the singlet and triplet states dissociating into the ground state $^2$S Rb atom
and the ground state $^2$S Ba$^+$ ion, electric transition dipole moments,
nonadiabatic and spin-orbit coupling matrix elements. Except for the spin-orbit
coupling and some nonadiabatic coupling matrix elements, all the results were
obtained with size-consitent methods based on the coupled cluster ansatz
including triple excitations. The nonrelativistic results are transformed to
the relativistic basis, thus allowing to judge the importance of the
relativistic effects, and to use our data in quantum dynamical
calculations both within the Hund's case $(a)$ and $(c)$.
Finally, we report single channel calculations of the elastic, spin-flip,
and charge transfer cross sections. This will allow us to obtain the
ratio of inelastic to elastic cross sections at various temperatures, and thus
give a first estimate of the efficiency of the sympathetic cooling of the barium
ion by collisions ultracold rubidium atoms.

The plan of this paper is as follows. In Sec.~\ref{sec2} we introduce the
theoretical models used in our calculations. We start
this section with a description
of the methods used in {\em ab initio} calculations of the Born-Oppenheimer potential
energy curves and electric transition dipole moments for (BaRb)$^+$.
We continue with the calculations of the nonadiabatic and spin-orbit
coupling matrix elements, and of the relativistic potentials.
The choice of
fixing the long-range coefficients at their {\em ab initio}
values is also addressed. The remaining part of this section is devoted
to the second step of the Born-Oppenheimer approximation, i.e. to dynamical
calculations of the elastic, spin-flip, and charge transfer scattering
cross sections. Numerical results are presented and
discussed in Sec.~\ref{sec3}. We start this section with the discussion of the
ground and excited states potentials. Next we turn to the nonadiabatic
coupling matrix elements and electric transition dipole moments. The effects of the
spin-orbit coupling on the potentials, transition moments, and nonadiabatic
coupling matrix elements are also addressed. Whenever possible, our
theoretical results are compared with the available {\em ab initio}
data \cite{Knecht:10,Knecht:10a}.
Once the results of the electronic structure calculations are presented
and discussed
we turn to the problem of producing cold barium ions. We present
the results for the elastic, spin-flip, and charge transfer cross sections
for both the nonrelativistic and relativistic potentials, and discuss the
efficiency of the sympathetic cooling leading to cold barium ions.
Finally, in Sec.~\ref{sec4} we conclude our paper.

\section{Computational details}
\label{sec2}
\subsection{Electronic structure calculations}
\label{sub1}
When dealing with collisions at ultra-low temperatures the accuracy of the potential
in the long range is very important. Therefore, the methods used in the
calculations should be size-consistent in order to ensure a proper dissociation
of the electronic state, and a proper long-range asymptotics of the potential.
In the present paper we adopt the computational scheme successfully applied
to the ground and excited states of the calcium dimer \cite{buss0,buss1,buss2,buss3,Koch:08}.
The potential energy
curves for the ground and excited states of the (BaRb)$^+$ molecular ion
have been obtained by a supermolecule method:
\be
V^{\rm ^{2S+1}|\Lambda|}(R)=
E_{\rm AB}^{\rm SM} -
E_{\rm A}^{\rm SM}-E_{\rm B}^{\rm SM}
\label{cccv}
\ee
where $E_{\rm AB}^{\rm SM}$ denotes
the energy of the dimer computed using the supermolecule method SM, and $E_{\rm X}^{\rm SM}$,
X=A or B, is the energy of the atom X.
For the ground state potential we used the coupled cluster method restricted to single, double, and
noniterative triple excitations, CCSD(T).
For the first triplet states of the $\Sigma$, $\Pi$, and $\Delta$ symmetries we
employed the restricted open-shell coupled cluster method restricted to single, double, and
noniterative triple excitations, RCCSD(T).
Calculations on all other excited states
employed the linear response theory (equation of motion) within the coupled-cluster singles,
doubles, and linear triples (LRCC3) framework~\cite{Jorgensen:90,Helgaker:97}.
Note that the second and higher triplet states and all excited singlet states
are open-shell systems that cannot be described with a single high-spin reference
function, so one has to resort to methods especially designed to describe open-shell
situations \cite{Jeziorski:10}.
The CCSD(T) and LRCC3 calculations were performed with the {\sc dalton}
program~\cite{dalton}, while RCCSD(T) calculations were done with the {\sc molpro}
suite of codes \cite{molpro}.
It is interesting to note at this point that even though (BaRb)$^+$ is effectively
a two-electron system, triple excitations are very important. A more detailed
discussion of this point will be presented in sec. \ref{sec3}.
In principle we could use the LRCC3 method to obtain all the excited state potentials
in any symmetry. However, the LRCC3 method is computationally more expensive than the
RCCSD(T) approach, so we decided to use the latter when possible. However, we have checked
for a few points that the RCCSD(T) and LRCC3 results for the $(1)^3\Sigma$, $(1)^3\Pi$, and
$(1)^3\Delta$ are very close.

The long-range asymptotics of the potentials is of primary importance for cold
collisions. Therefore, for each state we have computed the leading long-range
coefficients describing the electrostatic, induction, and dispersion interactions.
For the ground X$^1\Sigma$ state dissociating into Rb$^+$($^1$S)+Ba($^1$S)
and for the first excited singlet and triplet
states dissociating into the Ba$^+$($^2$S)+Rb($^2$S) the leading long-range
asymptotics at large internuclear distances $R$ reads:
\be
V^{\rm ^{2S+1}|\Lambda|}(R)= -\frac{C_4^{\rm ind}}{R^4} - \frac{C_6^{\rm ind}}{R^6}
- \frac{C_6^{\rm disp}}{R^6}+\cdots,
\label{lr1}
\ee
where the long-range coefficients $C_4^{\rm ind}$, $C_6^{\rm ind}$, and
$C_6^{\rm disp}$ are given by the standard expressions (see, e.g. Refs.
\cite{Jeziorski:94,Moszynski:08})
that can be derived from the multipole expansion of the interatomic interaction operator.
The long-range induction and dispersion coefficients were computed with the
recently introduced explicitly connected representation of the expectation value
and polarization propagator within the coupled cluster method \cite{Moszynski:93,Moszynski:05},
and the best approximation XCCSD4 proposed by Korona and collaborators \cite{Korona:06}.
For the singlet and triplet states dissociating into Ba$^+$($^2$S) ion and
Rb($^2$S) ion the induction coefficients were obtained from finite-field
RCCSD(T) calculations, while the dispersion coefficient from the sum-over-state
expression with the transition moments and excitation energies computed with the
multireference configuration interaction method limited to single and double
excitations (MRCI). Specifically, the transition moments and excitation energies
of the Ba$^+$($^2$S) ion were obtained in this way, while the Rb polarizability
at imaginary frequency was taken from highly accurate relativistic calculations
from the group of Derevianko \cite{Derevianko:10}.

For the molecular states of the (BaRb)$^+$ ion dissociating into Ba($^{1/3}$D)+Rb$^+$($^1$S)
the long-range asymptotics of the potentials is slightly more complicated and reads:
\be
V^{\rm ^{2S+1}|\Lambda|}(R)= \frac{C_3^{\rm elst}}{R^3} -
\frac{C_4^{\rm ind}}{R^4} + \frac{C_5^{\rm elst}}{R^5} - \frac{C_6^{\rm ind}}{R^6}
- \frac{C_6^{\rm disp}}{R^6}+\cdots,
\label{lr2}
\ee
where the new terms appearing in the expression above describe the long-range
charge-quadrupole ($C_3^{\rm elst}$) and charge-hexadecapole ($C_5^{\rm elst}$)
interactions. The mathematical expressions for the coefficients of Eq. (\ref{lr2})
are given by:
\be
C_3^{\rm elst} = (-1)^{2+\Lambda}
\pmatrix{ 2& 2 &2 \cr -\Lambda& 0&\Lambda \cr}
\langle^{1/3}{\rm D}||Q_2||^{1/3}{\rm D}\rangle,
\label{c3}
\ee
\be
C_4^{\rm ind} = \frac{1}{2}\left(\alpha_0 + \frac{3\Lambda^2-6}{6}\alpha_2\right)
\label{c4}
\ee
\be
C_5^{\rm elst} = (-1)^{2+\Lambda}
\pmatrix{ 2& 4 &2 \cr -\Lambda& 0&\Lambda \cr}
\langle^{1/3}{\rm D}||Q_4||^{1/3}{\rm D}\rangle,
\label{c5}
\ee
\be
C_6^{\rm ind} = \frac{1}{2}C_{zz,zz},
\label{c6ind}
\ee
\be
C_6^{\rm disp} = \frac{1}{2\pi}\int_0^\infty\alpha_0^{\rm Rb^+}(i\omega)
\left(6\alpha_0(i\omega)+3\frac{3\Lambda^2-6}{6}\alpha_2(i\omega)\right){\rm d}\omega.
\label{c6disp}
\ee
In these equations the expression in round brackets is a $3j$ symbol,
$\langle^{1/3}{\rm D}||Q_l||^{1/3}{\rm D}\rangle$, ($l=2$ or 4) denotes the reduced matrix
element of the quadrupole and hexadecapole moment, respectively,
$\alpha_0$  and $\alpha_2$ are the scalar  and tensor
components of the electric dipole polarizability tensor of the Ba atom in the
$^{1/3}$D state \cite{Kresnin:97}, while $C_{zz,zz}$ is the
$z$ component of the quadrupole polarizability
of Ba($^{1/3}$D). Finally, $\alpha_0^{\rm Rb^+}$ is the polarizability of
the rubidium ion Rb$^+$($^1$S). The electrostatic coefficients $C_3^{\rm elst}$
and $C_5^{\rm elst}$ for the $\Sigma$, $\Pi$, and $\Delta$ states are not
independent, and are connected one to the other by the following relations:
\be
C_3^{\rm elst}(\Sigma)=-C_3^{\rm elst}(\Delta)=2C_3^{\rm elst}(\Pi),
\; \; \; \; \;
C_5^{\rm elst}(\Sigma)=6C_5^{\rm elst}(\Delta)=-\frac{3}{2}C_5^{\rm elst}(\Pi).
\label{c3c5}
\ee
Note parenthetically that all coefficients
that lead to attractive interactions (induction and dispersion terms) are
assumed to be positive, while the electrostatic constants may result in
both attractive and repulsive interactions, so they enter with their true sign.
The values of the quadrupole moments, and scalar and tensor dipole
polarizabilities, as well as the components of the quadrupole polarizability
were obtained from finite-field LRCC3 calculations on the
$^{1/3}$D state of the atom. The hexadecapole moment of the singlet state
was obtained as an excited state expectation value
within the linear response CCSD formalism of Christiansen et al.
\cite{Helgaker:98}, while for the triplet state from the MRCI calculations.
The vector component of the dipole polarizability cannot be obtained from
finite-field calculations, so it was obtained from the sum-over-state
expression with the transition moments and excitation energies computed with the
MRCI method. All calculations of the long-range coefficients employed both
{\sc dalton} \cite{dalton} and {\sc molpro} \cite{molpro} suites of codes.

The transitions from the ground X$^1\Sigma$ state to the
$^1\Sigma$ and
$^1\Pi$ states are electric dipole allowed.
The transition dipole moments for the electric, $\mu_i$,
transitions were computed from the following expression
\cite{Bunker:98}:
\be
\mu_i=\langle\Psi_{\rm AB}^{\rm X}|r_i|\Psi_{\rm AB}^{\rm exc}\rangle,
\label{trandipel}
\ee
where
$r_i$ denotes the $i$th component of the position vector $\br$,
while
$\Psi_{\rm AB}^{\rm X}$ and $\Psi_{\rm AB}^{\rm exc}$ are the wave functions
for the ground and excited states, respectively.
Note that in Eq. (\ref{trandipel}) $i=x$ or $y$ corresponds to transitions
to $^1\Pi$ states, while $i=z$ corresponds to transitions to $^1\Sigma$
states.
In the present calculations the electric transition dipole moments
were computed as the first residue
of the LRCCSD linear response function with two electric, $\br$,
operators~\cite{Jorgensen:90}. In these calculations
we have used the {\sc dalton} program~\cite{dalton}.
We have evaluated the dependence of the transition dipole moments with
the internuclear distance
for the same set of distances as the excited state potential energy
curves.

As will be shown in the next section the electronic states of the
low lying excited states of the (BaRb)$^+$ molecular ion show strong
nonadiabatic couplings. Therefore, in this work we have computed the most
important radial
\be
R(R)=\langle(n)^{2S+1}|\Lambda||\frac{\partial}{\partial R}|(n')^{2S+1}|\Lambda|\rangle,
\label{rad}
\ee
and angular
\be
A(R)=\langle(n)^{2S+1}|\Lambda||L_+|(n')^{2S+1}|\Lambda'|\rangle,
\label{ang}
\ee
coupling matrix elements. In the above equations $\partial/\partial R$ and $L_+$
denote differentiation with respect to the internuclear distance and the electronic
angular momentum operator, respectively. Note that the radial operator couples
states of the same multiplicity and symmetry, while the electronic angular
momentum operator couples states with $\Lambda$ differing by one.
In the present calculations the angular coupling between the singlet states
was computed as the first residue
of the LRCCSD linear response function with two angular momentum
operators $\bL$~\cite{Jorgensen:90}. In these calculations
we have used the {\sc dalton} program~\cite{dalton}.
All other nonadiabatic couplings were obtained with the MRCI method and
the {\sc molpro} code \cite{molpro}.
We have evaluated the dependence of the nonadiabatic coupling matrix elements with
the internuclear distance
for the same set of distances as the excited state potential energy
curves.

Barium and rubidium are heavy systems, so
the electronic states of the (BaRb)$^+$ molecular ion are strongly mixed by the
spin-orbit (SO) interactions. Therefore, in any analysis of the collisional cross
sections between Ba$^+$ and Rb
the SO coupling and its dependence on the internuclear distance $R$
must be taken into account.
We have evaluated the spin-orbit coupling matrix elements
for the lowest dimer states
that couple to the 0$^+$ and 1 states of (BaRb)$^+$,
with the spin-orbit coupling operator $H_{\rm SO}$ defined within the Breit-Pauli
approximation \cite{Bethe:57}.
The spin-orbit coupling matrix elements
have been  computed within the MRCI framework
with the {\sc molpro} code~\cite{molpro}.
The full spin-orbit Hamiltonian has been used in the calculations, i.e.
both the one- and two-electron spin-orbit integrals were included.
Having the spin-orbit coupling matrix elements
at hand, we can build up the
matrices that will generate the potential energies  of the spin-orbit  states
that couple to  0$^+$ and 1 symmetry.
The matrix for the 1 states writes:
\begin{equation}
\scriptsize{
\left(
\begin{array}{ccccc}
V^{(1)^{1}\Pi} &  \langle(1)^{1}\Pi |H_{\rm SO}|(1)^{3}\Sigma \rangle &  \langle(1)^{1}\Pi |H_{\rm SO}|(1)^{3}\Delta \rangle &  \langle(1)^{1}\Pi |H_{\rm SO}|(2)^{3}\Sigma \rangle & \langle(1)^{1}\Pi |H_{\rm SO}|(1)^{3}\Pi \rangle \\
 \langle (1)^{3}\Sigma |H_{\rm SO}|(1)^{1}\Pi  \rangle & V^{(1)^{3}\Sigma} & 0 & 0 &  \langle (1)^{3}\Sigma  |H_{\rm SO}|(1)^{3}\Pi  \rangle \\
 \langle(1)^{3}\Delta|H_{\rm SO}|(1)^{1}\Pi \rangle & 0 & V^{(1)^{3}\Delta} - \langle(1)^{3}\Delta|H_{\rm SO}|(1)^{3}\Delta \rangle  & 0 &  \langle(1)^{3}\Delta  |H_{\rm SO}|(1)^{3}\Pi  \rangle \\
 \langle (2)^{3}\Sigma  |H_{\rm SO}|(1)^{1}\Pi  \rangle & 0 & 0 &  V^{(2)^{3}\Sigma} &  \langle (2)^{3}\Sigma  |H_{\rm SO}|(1)^{3}\Pi  \rangle \\
\langle(1)^{3}\Pi  |H_{\rm SO}|(1)^{1}\Pi  \rangle & \langle (1)^{3}\Pi  |H_{\rm SO}| (1)^{3}\Sigma  \rangle & \langle (1)^{3}\Pi |  |H_{\rm SO}|(1)^{3}\Delta  \rangle & \langle (1)^{3}\Pi |H_{\rm SO}|  (2)^{3}\Sigma  \rangle & V^{(1)^{3}\Pi}
\end{array}
\right)}
\label{SO1}
\end{equation}
while the matrix for the 0$^+$ states is given by:
\begin{equation}
\footnotesize{
\left(
\begin{array}{cccc}
 V^{(1)^{3}\Pi}- \langle (1)^{3}\Pi|H_{\rm SO}|(1)^{3}\Pi \rangle  &  \langle (1)^{3}\Pi |H_{\rm SO}|{\rm X}^{1}\Sigma \rangle & \langle (1)^{3}\Pi |H_{\rm SO}|(2)^{1}\Sigma \rangle &\langle (1)^{3}\Pi |H_{\rm SO}|(1)^{1}\Sigma \rangle \\
\langle {\rm X}^{1}\Sigma  |H_{\rm SO}|(1) ^{3}\Pi \rangle & V^{{\rm X}^{1}\Sigma}  & 0 & 0 \\
\langle (2)^{1}\Sigma  |H_{\rm SO}| (1)^{3}\Pi \rangle & 0&  V^{(2)^{1}\Sigma} &0\\
\langle (3)^{1}\Sigma  |H_{\rm SO}| (1)^{3}\Pi \rangle & 0&0 &  V^{(3)^{1}\Sigma} \\
\end{array}
\right)}
\label{SO0}
\end{equation}
Diagonalization of these matrices gives the spin-orbit coupled potential
energy curves for the $1$ and $0^+$ states, respectively. Note that
all potentials in the matrices (\ref{SO1}) and (\ref{SO0}) were
taken from CCSD(T), RCCSD(T), and LRCC3 calculations. Only the
diagonal and nondiagonal spin-orbit coupling matrix elements were
obtained with the MRCI method.
Once the eigenvectors of these matrices are available, one can easily
get the electric dipole transition moments and the nonadiabatic coupling
matrix elements between the relativistic states.

Finally, the long-range coefficients corresponding to the relativistic
potentials were obtained by diagonalizing the matrices of Eq. (\ref{SO1})
and (\ref{SO0}) with the potentials expanded according to Eqs. (\ref{lr1})--(\ref{lr2})
and the spin-orbit coupling matrix elements fixed at the atomic values. Note that
unlike in the case of resonant interactions between like atoms \cite{skomo}, the SO coupling
does not change the leading power in the asymptotic expansion of the
interaction energy, but only changes the numerical values of the coefficients.
One should also note that the atomic SO coupling does not change in our model the long-range
coefficients for the (1)$^1\Sigma$ and (2)$^1\Sigma$ states
due to the different dissociation limits:
Ba($^{1/3}$D)+Rb$^+$($^1$S) versus Ba$^+$($^2$S)+Rb($^2$S).

In order
to mimic the scalar relativistic effects some electrons were described
by pseudopotentials. For Ba and Rb we took the ECP46MDF
\cite{Stoll:06} and ECP28MDF \cite{Stoll:05}
pseudopotentials, respectively, from the Stuttgart library.
For both barium and rubidium we used the $spdfg$ quality basis sets
suggested in Refs. \cite{Stoll:06,Stoll:05}.
The full basis of the dimer was used in the supermolecule
calculations and the Boys and Bernardi scheme was used
to correct for the basis-set superposition error \cite{Boys:71}.

It should be stressed at this point that the {\em ab initio} results reported in the
present paper obtained by the {\em ab initio} methods described above
will allow to perform dynamical calculations of the cross sections
in the nonadiabatic, multichannel regime, both in the Hund's case $(a)$ (nonrelativistic
states, SO and nonadiabatic couplings, and transition moments) and $(c)$
(relativistic potentials, nonadiabatic couplings, and transition moments), cf.
Ref. \cite{Carrington}.

Finally, to conclude this section we would like to emphasize that almost all {\em ab initio}
results were obtained with the most advanced size-consistent methods of quantum
chemistry: CCSD(T), RCCSD(T), and LRCC3. Only the SO coupling matrix elements
and nonadiabatic matrix elements were obtained with the MRCI method which is
not size consistent. Fortunately enough, all of the couplings are important
in the region of the curve crossings or avoided crossings and vanish at large
distances, so the effect of the size-inconsistency of MRCI on our results
should not be drammatic.

\subsection{Dynamical calculations}
\label{sub2}
In collisions of the Ba$^+$($^2$S) ion with Rb($^2$S) atom we have basically
three types of processes: elastic scattering in the singlet and triplet
potentials, spin-flip (spin-exchange) process, and the inelastic radiative
charge transfer from the singlet and triplet manifolds of Ba$^+$Rb to the ground
state of BaRb$^+$.
In the present paper we restrict ourselves to single channel calculations.
A more detailed multichannel treatment will be presented elsewhere.

To compute the elastic cross sections we need to solve the radial Schr\"odinger
equation for the relative motion of the Ba$^+$ ion and Rb atom at an energy $E$:
\be
\left(\frac{d^2}{dR^2} - \frac{2\mu}{\hbar^2} V(R)
-\frac{J(J+1)}{R^2}+
\frac{2\mu E}{\hbar^2}\right)\Psi_{EJ}(R)=0,
\label{sch1}
\ee
subject to the following normalization conditions:
\be
\int_0^\infty\Psi_{E_1J}^\star(R)\Psi_{E_2J}(R){\rm d}R=
\delta(E_1-E_2),
\label{norm1}
\ee
where $\Psi_{EJ}(R)$ is the scattering wave function, $\mu$
is the reduced mass of the (BaRb)$^+$ ion, and $V(R)$ stands
for the interaction potential of Ba$^+$($^2$S) with Rb($^2$S) in the
singlet or triplet manifold.
Note that the normalization condition of Eq. (\ref{norm1}) is equivalent
to the following large-$R$ behavior of the wave function $\Psi_{EJ}(R)$,
\be
\Psi_{EJ}(R)\sim \left(\frac{2\mu}{\pi\hbar^2 k}\right)^{1/2}
\sin\left(kR-\frac{J\pi}{2}+\delta_J(E)\right),
\label{norm3}
\ee
where $\delta_J(E)$ denote the phase shift corresponding to the $J$ partial wave, and
the wave vector $k$ is given by the standard expression,
$E=\frac{\hbar^2k^2}{2\mu}$.
Equation (\ref{sch1}) subject to the normalization condition (\ref{norm1})
allows us to compute the cross sections for the elastic and spin-flip
collisions from the standard expressions:
\be
\sigma_{\rm el}^{\rm s}(E) = \frac{4\pi}{k^2}\sum_{J=0}^\infty (2J+1)\sin^2\delta_J^{\rm s}(E),
\; \; \; \; \;
\sigma_{\rm el}^{\rm t}(E) = \frac{4\pi}{k^2}\sum_{J=0}^\infty (2J+1)\sin^2\delta_J^{\rm t}(E),
\label{el}
\ee
\be
\sigma_{\rm sf}(E) = \frac{4\pi}{k^2}\sum_{J=0}^\infty (2J+1)\sin^2\left(\delta_J^{\rm s}(E)
-\delta_J^{\rm t}(E)\right),
\label{sf}
\ee
where the superscripts ``s'' and ``t'' on $\sigma_{\rm el}$
and $\delta_J$ pertain to the singlet and triplet potentials,
respectively. Note that an exact description of the spin-flip process would require at
least two coupled channels, so the expression (\ref{sf}) is only approximate
\cite{Dalgarno:65}. It is derived under assumption that the hyperfine splittings are much smaller than the collision energy.
However, it was shown to work relatively well, even at low energies \cite{Stoof:88,Cote:98}.

Theoretical description of the charge transfer process between the atom and the ion
is somewhat more ellaborate. To the first-order of perturbation theory the radiative
charge transfer can be described by the following Fermi golden type expression
\cite{Dalgarno:88}:
\be
\sigma_{\rm ct}(E)=\frac{4\pi^2\hbar}{k^2}A(E),
\label{ct}
\ee
where the Einstein coefficient $A(E)$ is given by:
\begin{eqnarray}
A(E) && =
\frac{4\alpha^3}{3e^4\hbar^2}
\sum_{J'=0}^\infty\sum_{J''=J'\pm 1}(2J'+1)[
\int_0^\infty \varepsilon^3 H_{J'}|\langle\Psi_{E'J'}|\mu(R)|\Psi_{E''J''}\rangle|^2
{\rm d}\varepsilon
\cr
&&
+
\sum_{v''}
H_{J'}
(E_{v''J''}-E')^3
|\langle\Psi_{E'J'}|\mu(R)|\Psi_{v''J''}\rangle|^2],
\label{Acoef}
\end{eqnarray}
where the primed and double primed quantities pertain to the excited and
ground state potentials, respectively, $\mu(R)$ is the transition moment
from the ground to the excited electronic state, $\alpha=1/137.035999679(94)$
is the fine structure constant, $e$ is the electron charge, and
the H\"ohn-London factor $H_{J'}$ is equal to $(J'+1)/(2J'+1)$ for
the $P$ branch ($J'=J''-1$), and to $J'/(2J'+1)$ for the $R$ branch
($J'=J''+1$). Here, $\varepsilon$
stands for the energy difference
\be
\varepsilon=E'' - E' + \Delta_{\rm IP},
\label{DIP}
\ee
where $\Delta_{\rm IP}$ is the difference of the ionization potentials.
The scattering wave functions appearing in the expression (\ref{Acoef})
are solutions of Eq. (\ref{sch1}), while the
bound-state wave functions fulfill
the following Schr\"odinger equation:
\be
\left(\frac{d^2}{dR^2} - \frac{2\mu}{\hbar^2} V(R)
-\frac{J''(J''+1)}{R^2}+
\frac{2\mu E_{v''J''}}{\hbar^2}\right)\Psi_{v''J''}(R)=0,
\label{schr2}
\ee
subject to the following normalization conditions:
\be
\int_0^\infty\Psi_{v_1J}^\star(R)\Psi_{v_2J}(R){\rm d}R=
\delta_{v_1 v_2},
\label{norm2}
\ee
where $V(R)$ stands for the ground state potential of the (BaRb)$^+$ molecular
ion.

A significantly simpler approach proposed in Ref. \cite{Julienne:84} approximates
the sum over all continuum and bound states in Eq. (\ref{Acoef}) with a simple
average of a space-varying Einstein coefficient over the initial scattering wave
function $\Psi_{E'J'}$:
\be
A(E) = \sum_{J'=0}^\infty (2J'+1) \langle\Psi_{E'J'}|\bar{A}(R)|\Psi_{E'J'}\rangle,
\label{Aapprox}
\ee
where
\be
\bar{A}(R) = \frac{\alpha^3}{3\hbar e^6}(\delta V)^3(R)\mu^2(R),
\label{Abar}
\ee
and $\delta V(R)$ is the difference between the excited and ground state
potentials.

\section{Numerical results and discussion}
\label{sec3}
\subsection{Nonrelativistic potential energy curves and spectroscopic characteristics
of the ground and excited states}
\label{sub3}
\begin{figure}
\begin{center}
\begin{minipage}{\textwidth}
\epsfxsize=14cm \epsffile{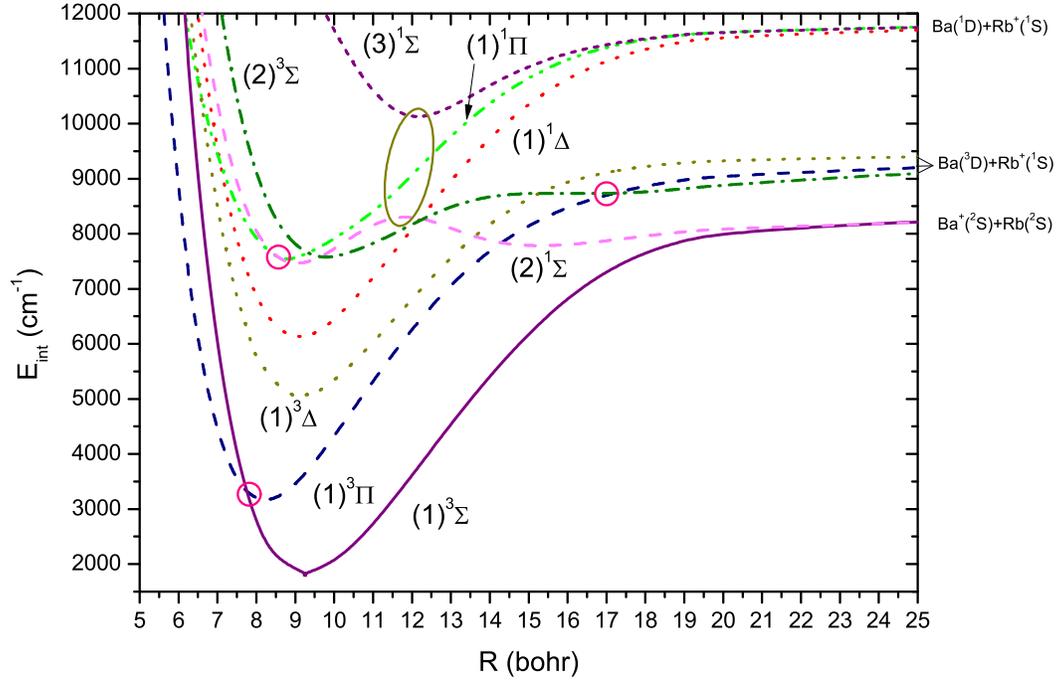}
\end{minipage}
\end{center}
\caption{Nonrelativistic potential energy
curves for the excited states of the (BaRb)$^+$ molecular ion.}
\label{fig1}
\end{figure}

Calculations were done for the ground state and first eight
(four singlet and four triplet) excited states of (BaRb)$^+$.
Two states dissociate into Ba$^+$($^2$S)+Rb($^2$S),
three states into   Ba($^3$D) +  Rb$^+$($^1$S), and three states into
Ba($^1$D) +  Rb$^+$($^1$S).
The  potential energies were calculated for twenty interatomic distances
$R$ ranging from 4 to 50 bohr for each potential curve.
The potential curves  are plotted in Fig.~\ref{fig1}, while the
spectroscopic characteristics of these
states are reported in Table \ref{tab1}. The ground state is absent on figures due to its regular behaviour and single minimum, and in order to increase the visibility of the other states. The separated atoms energy for each state was set equal to the experimental
value. Numerical values  of the potentials are available from the authors on request.

Before going on with the discussion of the potentials let us note that the atomic
excitation energies obtained from the LRCC3 calculations are very accurate.
Our predicted position of the nonrelativistic $^3$D state of barium is 9422 cm$^{-1}$,
to be compared with the experimental value of 9357 cm$^{-1}$ \cite{nist} deduced from
the positions of the states in the D multiplet and the Land\'e rule.
For the $^1$D state of Ba we obtain 11907 cm$^{-1}$, in a relatively good
agreement with the experimental value of 11395 cm$^{-1}$ \cite{nist}. It is worth
noting that the present results for the atomic excitation energies are as accurate
as the results of fully relativistic atomic calculations of Kozlov and Porsev
\cite{Porsev:99}, and more accurate than the data obtained from fully
relativistic Dirac-Coulomb calculations \cite{Knecht:10a}.
To further assess the quality of the methods, basis-sets and pseudopotentials employed in the
present paper we have computed the static polarizabilities of the ground state of Ba
atom, of the
ground state of the Ba$^+$ ion, and the scalar and tensor components of the polarizability
of Ba($^1$D). The present polarizabiity of the ground state of the barium atom is
272.5 a.u. The experimental value is $268\pm 22$ \cite{Bederson:74}, while the best
theoretical result of Kozlov and Porsev \cite{Porsev:99} is 264 a.u.. Also the static
polarizability of the Ba$^+$ ion, 132 a.u., is in a fairly good agreement with the
result of fully relativistic calculations of Ref. \cite{Safronova:08}, 124.15 a.u.,
of Ref. \cite{Mukherjee:09}, 124.26 a.u., and with the most recent experimental result, 123.88(5) a.u.
\cite{Lundeen:07}. For the $^1$D state
we get the scalar and tensor polarizabilities of 289 and 73 a.u., in a fair agreement
with the results of Ref. \cite{Porsev:99}, 266 and 81 a.u., respectively.
The methods employed in the present paper do not allow for a consistent calculation
of the dissociation limit $\Delta_{\rm IP}$
of the $(2)^1\Sigma$ and $(1)^3\Sigma$ states corresponding
to Ba$^+$($^2$S)+Rb($^2$S) separated atoms, since the ground state calculations employed
CCSD(T) and the calculations on the singlet and triplet excited states the LRCC3
and RCCSD(T) methods, respectively. However, we can estimate $\Delta_{\rm IP}$
from the energy of the $(2)^1\Sigma$ and $(1)^3\Sigma$ at $R=50$ bohr. Both the
LRCC3 calculation on the singlet state and RCCSD(T) calculation on the triplet
state give $\Delta_{\rm IP}=8097$ cm$^{-1}$ in a good agreement with the
experimental value of 8344 cm$^{-1}$ \cite{nist}. The
fully relativistic result of Ref. \cite{Knecht:10a}, 8065 cm$^{-1}$, is very close
to our value.

The ground X$^1\Sigma$ state of the (BaRb)$^+$ molecular ion is a strongly bound
state with the binding energy of 5136 cm$^{-1}$. The minimum on the potential
energy curve for this state appears at a relatively large distance $R_e=8.67$
bohr. The origins of the binding can be explained by using the symmetry-adapted
perturbation theory of intermolecular forces (SAPT) \cite{Jeziorski:94,Moszynski:08}.
As could be guessed, the interaction energy at the minimum results from a
subtle balance of the induction attraction and exchange-repulsion. The
induction energy is huge, --14499 cm$^{-1}$, but is strongly quenched by the
exchange-repulsion term, 10068 cm$^{-1}$. The electrostatic contribution due
to the charge overlap of the unperturbed electron clouds of Ba and Rb$^+$,
--554 cm$^{-1}$, and dispersion term, --598 cm$^{-1}$ is
of minor importance, again in agreement with our intuition.

An inspection of Fig. \ref{fig1} shows that the
potential energy curves for the excited states of the (BaRb)$^+$ molecular ion are
smooth with well defined minima. The potential energy curves of the (2) and
(3)$^1\Sigma$ show an avoided crossing. The potential energy curves of the
(2)$^1\Sigma$  and (2)$^3\Sigma$  states exhibit a double minimum structure.
The double minimum on the potential energy curve of the (2)$^1\Sigma$ state
is due to the interaction with (3)$^1\Sigma$ state. Other double minimum
structure can be explained from the long-range theory and will be discussed
below.
Some potential energy curves show maxima. These are due to the
first-order electrostatic interactions in the long range, and will also be
discussed in more details in sec. \ref{sub4}.
Except for the shallow double minima structure of the (2)$^1\Sigma$ and (2)$^3\Sigma$ states,
and a shallow (3)$^1\Sigma$ state,
all other excited states of the (BaRb)$^+$ molecular ion are strongly bound with binding
energies $D_e$ ranging from
approximately 4380 cm$^{-1}$ for the (1)$^3\Delta$ state up to
6301 cm$^{-1}$ for the (1)$^3\Pi$ state. The (2)$^1\Sigma$ state, important from
the experimental point of view, has two minima at $R_e$= 9.02 and 15.19 bohr.
The depths of these minima are 911 and 576 cm$^{-1}$, while the barrier separating
them is of 30 cm$^{-1}$ suggesting that the tunneling between the two wells will
be very fast. The (2)$^3\Sigma$ state also shows two minima at $R_e$ 9.82 and 16.78 bohr of
1874 and 697 cm$^{-1}$, respectively, separated by a barrier of 681 cm$^{-1}$.

\begin{figure}
\begin{center}
\begin{minipage}{\textwidth}
\epsfxsize=14cm \epsffile{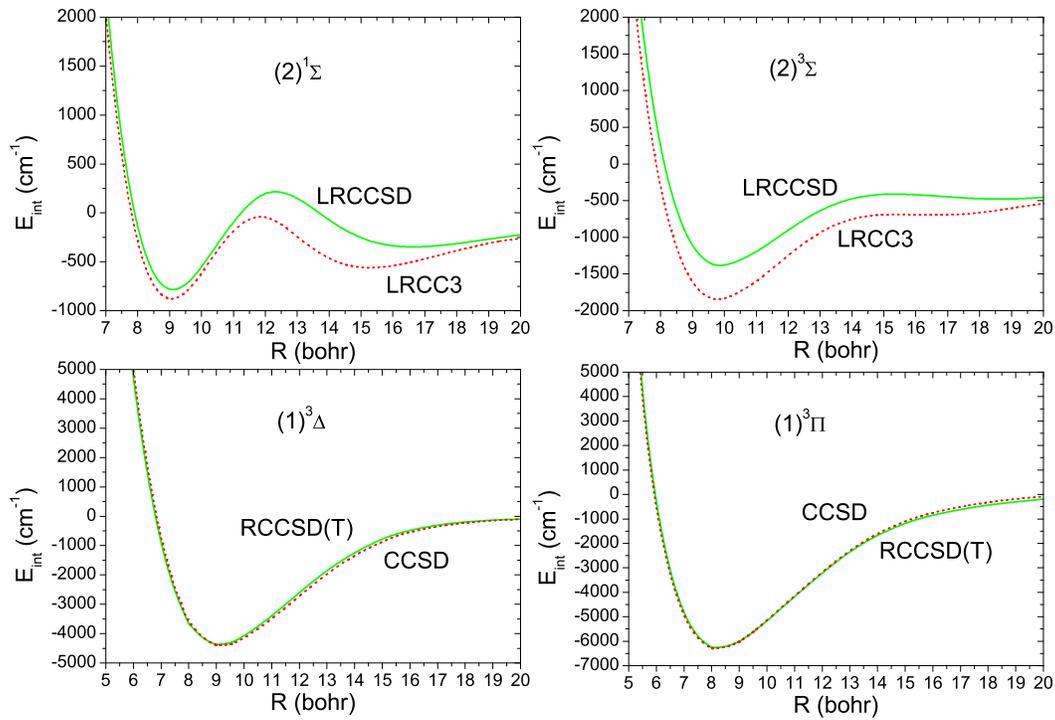}
\end{minipage}
\end{center}
\caption{Comparison of the LRCC3 with CCSD, and RCCSD(T) with LRCCSD
potential energy
curves for selected states of the (BaRb)$^+$ molecular ion.}
\label{fig1a}
\end{figure}

Before comparing our results with the {\em ab initio} data reported in Refs.
\cite{Knecht:10,Knecht:10a} let us stress the importance of the triple
excitations in the wave functions for some states.
This point is illustrated in
Fig. \ref{fig1a}, where LRCCSD versus LRCC3 and CCSD versus CCSD(T) potential
energy curves are plotted for selected states.
An inspection of these figures shows that the contribution of the
triple excitations
is important for the excited (2)$^1\Sigma$ and (2)$^3\Sigma$ states,
and relatively unimportant for the
(1)$^3\Delta$ and for the (1)$^3\Pi$ states.
Not shown on these figures are the (1)$^3\Sigma$, (1)$^1\Pi$, and (1)$^1\Delta$ states.
For these states we find that the potential for the (1)$^3\Sigma$ state
is relatively unaffected by the triple excitations, while the two other
potential energy curves show large differences depending whether the $T_3$
cluster operator is included or not in the wave functions. These results
strongly suggest that the CCSD or LRCCSD method works for those states
that can be described by a single reference determinant. For open-shell
states, i.e. all singlet states and the (2)$^3\Sigma$ state, the effect
of the triple excitations is large and changes both the well depths
and the barriers.
In principle, the differences could be due to CCSD(T) vs. LRCC3 methods
used in the calculations. However, as we already stated in sec. \ref{sec2},
we have checked
for a few points that the RCCSD(T) and LRCC3 results for the $(1)^3\Sigma$, $(1)^3\Pi$, and
$(1)^3\Delta$ are very close.
To get a better understanding of the importance of the triple excitations
we analysed the energy gaps between the
lowest electron configuration for a given state and the lowest
triply excited configuration. It turned out that for the effectively high-spin
 states this energy difference was high, while for the
manifestly states the opposite was true. Using simple
perturbation theory arguments one can deduce that the
correlation due to the triple excitations will be important for states
with a small energy gap between the lowest and triply
excited configurations, and much less
important for states with large energy difference.

Let us compare our results with other available {\em ab initio} data \cite{Knecht:10}.
The spectroscopic constants are listed in Table~\ref{tab1} and compared with
the results of Ref. \cite{Knecht:10}. Unfortunately, Knecht et al. \cite{Knecht:10}
did not report the binding energies $D_e$ of the molecular states, but only the
values of the electronic term values $T_e$ taking the minimum of the ground state
as zero of energy. An inspection of Table \ref{tab1} shows that the agreement
with the data of Knecht et al. \cite{Knecht:10}
is relatively good, given the fact that their
results were obtained using the internally contracted multireference
configuration singles and doubles
method based on a CASSCF reference function.
For most of the states the computed electronic terms agree within
a few hundred cm$^{-1}$. For the high-spin states the differences
in the positions of the minima are 0.1 bohr at worst, while the
electronic terms differ by 300 to 600 cm$^{-1}$. For the open-shell
states, where the triple excitations in the wave function are
important, the differences in the positions and well depths are
more important. The most striking difference between the present
results and the data of Ref. \cite{Knecht:10} is the (1)$^1\Pi$ state.
Here, the difference in the position of the minimum is 0.6 bohr,
and the difference in $T_e$ is as much as 3300 cm$^{-1}$.
The double minimum structure of the (2)$^1\Sigma$ and (2)$^3\Sigma$
states was not reproduced by MRCI calculations, but the barriers
and the avoided crossing between the (2)$^1\Sigma$ and (3)$^1\Sigma$
states are reproduced. In general, the agreement is only qualitative.
To end this paragraph we would like to say that the results of
fully relativistic Dirac-Coulomb calculations published by the
same Authors in Ref. \cite{Knecht:10a} are in a much better
agreement with the present data, cf. sec. \ref{sub5}.

\subsection{Nonadiabatic coupling matrix elements
and electric transition dipole moments from the ground X$^1\Sigma$ state}
\label{sub4}
\begin{figure}
\begin{center}
\begin{minipage}{\textwidth}
\epsfxsize=14cm \epsffile{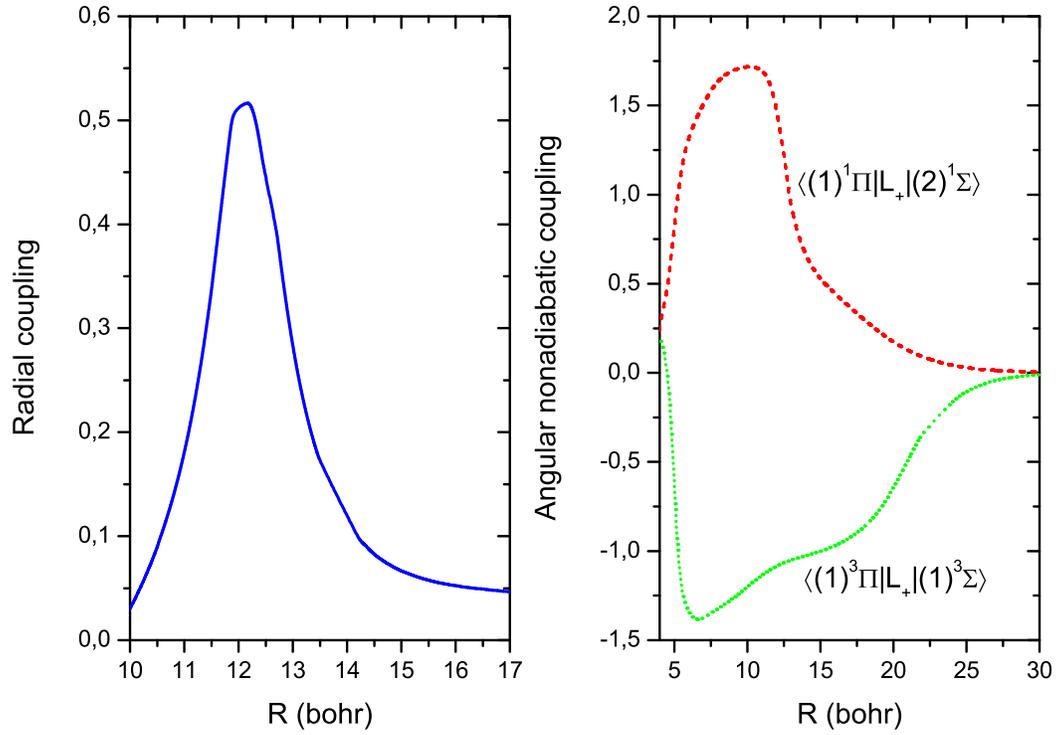}
\end{minipage}
\end{center}
\caption{Nonadiabatic coupling matrix elements of the nonrelativistic electronic states
of the (BaRb)$^+$ molecular ion. The left and right pannels correspond to the radial and
angular couplings, respectively.}
\label{fig2}
\end{figure}

\begin{figure}
\begin{center}
\begin{minipage}{\textwidth}
\epsfxsize=14cm \epsffile{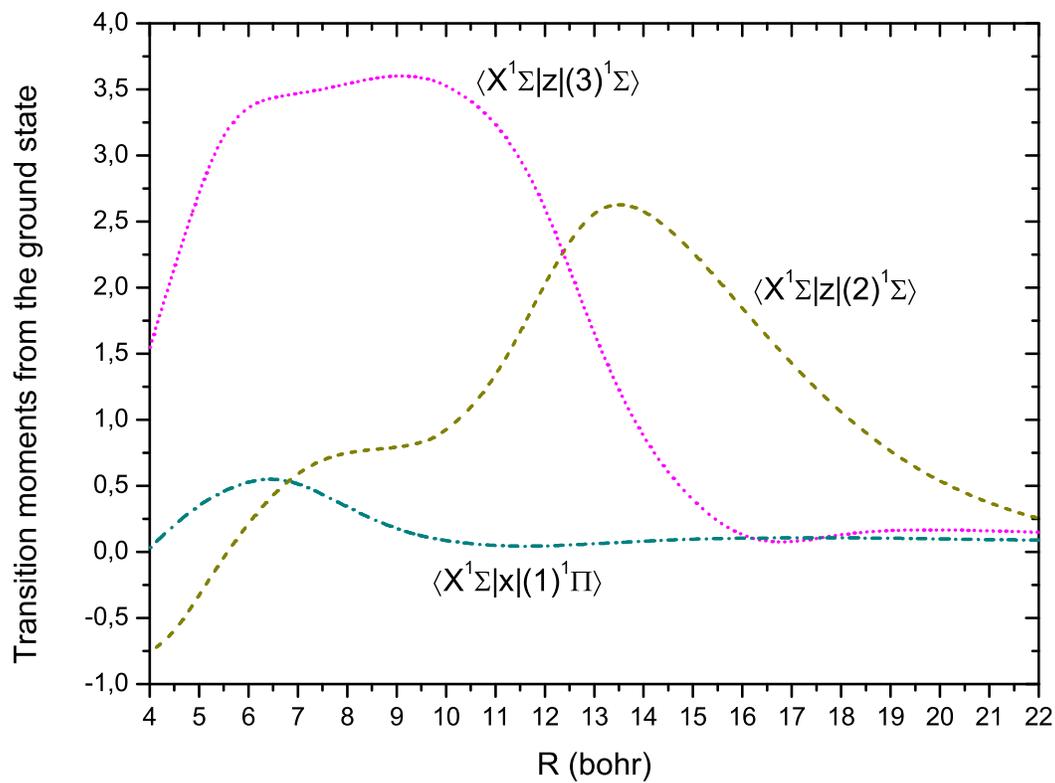}
\end{minipage}
\end{center}
\caption{Electric transition dipole moments from the ground X$^1\Sigma$ state to
the $^1\Sigma$ and $^1\Pi$  states of (BaRb)$^+$.}
\label{fig3}
\end{figure}

The most important nonadiabatic coupling matrix elements between the
excited states of the (BaRb)$^+$ molecular ion are reported on
Fig. \ref{fig2}. The regions where these couplings could possibly
be important are indicated on Fig. \ref{fig1}.
The left pannel on this figure shows the radial
coupling between the $(2)^1\Sigma$ and $(3)^1\Sigma$ states.
The potential energy curves for these states reveal an avoided
crossing shown on Fig. \ref{fig1} at $R\approx 12$ bohr. An
inspection of Fig. \ref{fig2} shows that the maximum of the
radial coupling between these two states corresponds to this
distance. In general, the radial coupling as a function of the
distance $R$ is small, and rather localized around  the
point of the avoided crossing. The angular coupling matrix
elements reported on the right pannel of Fig. \ref{fig3} show
more variations with $R$. The angular coupling between the (1)$^1\Pi$ and
(2)$^1\Sigma$ states has a broad maximum around $R\approx 11$ bohr,
and this distance roughly corresponds to the crossing of the (1)$^1\Pi$ and
(2)$^1\Sigma$ potential energy curves.
The angular coupling between the (1)$^3\Pi$ and
(1)$^3\Sigma$ states shows a broad minimum at $R\approx 7$ bohr,
and again this distance roughly corresponds to the crossing of the (1)$^3\Pi$ and
(1)$^3\Sigma$ potential energy curves.
The last angular coupling that may influence the dynamics of the (BaRb)$^+$
molecular ion corresponds to the (2)$^3\Sigma$ and (1)$^3\Pi$ states.
Here, the $R$ dependence of the angular coupling is quite different,
but the largest variations correspond again to the point, where the
two curves cross, $R\approx 12$ bohr. Note parenthetically that
radial coupling tends to zero as $R^{-7}$, the angular couplings
between the (2)$^3\Sigma$ and (1)$^3\Pi$ states tends to a constant
value, $\langle^3{\rm D}(M_L=1)|L_+|^3{\rm D}(M_L=0)\rangle=\sqrt{3}$, while
the coupling between the (1)$^1\Pi$ and (2)$^1\Sigma$ decays
exponentially at large $R$. This
unusual exponential decay is due to the different dissociation
limits of the ground X$^1\Sigma$ and excited $(2)^1\Sigma$ states:
Ba($^1$S)+Rb$^+$($^1$S) versus Ba$^+$($^2$S)+Rb($^2$S).
It is gratifying to note that the MRCI method used in the calculations
of the radial coupling and angular couplings between the
triplet states quite precisely located the regions of the avoided
crossing and curve crossing despite the fact the latter were
determined from the RCCSD(T) and LRCC3 calculations. This suggests
that the computed nonadiabatic coupling matrix elements are
reliable, at least around of the crossings.

The electric dipole transition moments  between the ground state and the
three excited electric dipole-allowed states,
two $^1\Sigma$ and  one $^1\Pi$,
are plotted in Fig.~\ref{fig3} as
functions of the interatomic distance $R$.
The calculated electric transition moments show a strong dependence
on the internuclear distance $R$.
For the transition moments to the excited states of the $\Sigma$
symmetry the curves show broad maxima around the positions of the
depths on the potential energy curves. The transition moment to
the $\Pi$ state is very small, suggesting that this state will be
of minor importance in the dynamics of the (BaRb)$^+$ ion.
At large interatomic distances
the transition moments to the $(3)^1\Sigma$ and
$(1)^1\Pi$ states tend to zero
as $\mu_4R^{-4}$, while the transition moment to the $(2)^1\Sigma$ state
decays exponentially with the internuclear distance $R$.
This $R^{-4}$ dependence can be derived fron the multipole
expansion of the wave functions of the $(3)^1\Sigma$, $(1)^1\Pi$,
and X$^1\Sigma$ states. The expressions for the leading long-range
coefficient $\mu_4$ of the electric transition dipole moments read:
\be
\mu_4=A_{|\Lambda|}\alpha_0^{\rm Rb^+}\langle^1{\rm S}||Q_2||^1{\rm D}\rangle,
\label{mu4}
\ee
where $A_0=3/2\sqrt{5}$ and $A_1=A_0/\sqrt{3}$.

\subsection{Spin-orbit coupling and relativistic potential energy curves, nonadiabatc
coupling, and electric transition dipole moments}
\label{sub5}
\begin{figure}
\begin{center}
\begin{minipage}{\textwidth}
\epsfxsize=14cm \epsffile{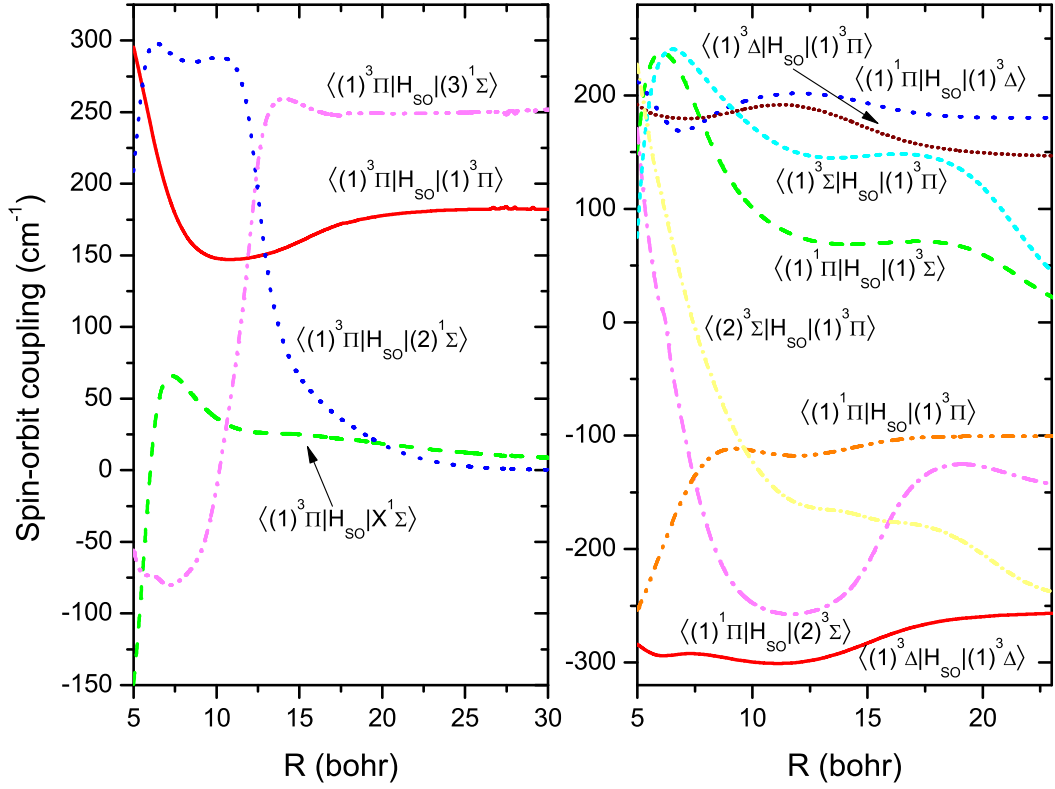}
\end{minipage}
\end{center}
\caption{Matrix elements of the spin-orbit coupling for the electronic states
of (BaRb)$^+$. The left and right pannels correspond to $0^+$ and 1 states, respectively.}
\label{fig4}
\end{figure}

A large number of spin-orbit interactions couple the dimer states of
the (BaRb)$^+$ molecular ion. In Fig. \ref{fig4} we report the $R$ dependence of
the spin-orbit coupling matrix elements that couple to the
1 and 0$^+$  states. These states are most interesting for  the collisional dynamics
of (BaRb)$^+$. Similar results can easily be obtained for the
0$^-$, 2, and 3 states. An inspection of Fig. \ref{fig4} shows
that the SO coupling matrix elements have relatively large variations
at small internuclear distances and tend to zero or to the atomic values at large $R$.
All diagonal matrix elements of the spin-orbit Hamiltonian are
lower than the atomic $1^3$D spin-orbit constant of barium
($\approx$ 150 cm$^{-1}$),
while the largest nondiagonal couplings are observed  for the
pairs of states associated with
the crossing of the corresponding potential energy curves.
The accuracy of the atomic SO couplings can be judged by comparison of the computed
and observed positions of the energy levels in the $^3$D multiplet.
The calculated
energies of the $^3$D$_1$, $^3$D$_2$, and $^3$D$_3$ states are 9035, 9254, and 9680
cm$^{-1}$, and are in a very good agreement
with the experimental values of 9034, 9216, and 9597 cm$^{-1}$
\cite{nist}.
It is worth noting that some of the couplings vanish at large distances
due to the different dissociation limits:
Ba+Rb$^+$ versus Ba$^+$+Rb.
This means that the neglect of the $R$ dependence of the spin-orbit
matrix elements would lead to wrong relativistic potentials since some of important
couplings would be neglected.
As an example we can cite the coupling of the (1)$^3\Sigma$
and (1)$^3\Pi$ states. The asymptotic value is zero due to
the different dissociation
limits of these states:
Ba($^3$D)+Rb$^+$($^1$S) versus Ba$^+$($^2$S)+Rb($^2$S).
Therefore, when approximating in Eq. (\ref{SO1}) all SO couplings by the corresponding
atomic values one would obtain a completely wrong matrix
of the spin-orbit Hamiltonian for the 1 states since
in the atomic limit $\langle (1)^{3}\Sigma  |H_{\rm SO}|(1)^{3}\Pi\rangle$=0.
In particular, in the atomic approximation the (1)$^3\Sigma$ state would remain
unchanged, and as it will be shown below it changes quite a lot at small
internuclear distances.
The same is true for the (2)$^1\Sigma$ state.

\begin{figure}
\begin{center}
\begin{minipage}{\textwidth}
\epsfxsize=14cm \epsffile{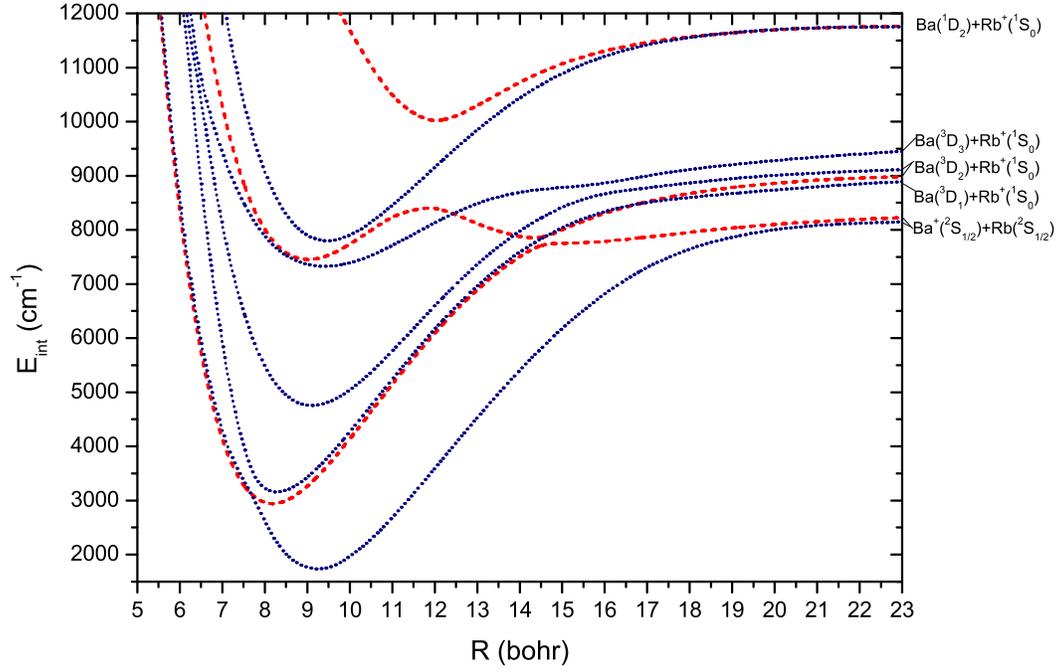}
\end{minipage}
\end{center}
\caption{Relativistic potential energy curves for the excited states of the (BaRb)$^+$ molecular ion.
Red dashed lines correspond to the $0^+$ states and the blue dotted lines to the $1$ states.}
\label{fig5}
\end{figure}

The diagonalization of the spin-orbit Hamiltonian matrices, Eqs.
(\ref{SO1}) and (\ref{SO0}), gives
the potential energy curves of the states that couple
to 1 or $0^+$ symmetry. The corresponding curves
for the $0^+$ and 1 symmetries are reported on
Fig. \ref{fig5}.  An inspection of Figs. \ref{fig1} and \ref{fig5} shows that
the crossings of the diabatic (nonrelativistic) states are
transformed into  avoided crossings on the spin-orbit
coupled relativistic curves. Obviously,
the inclusion of the spin-orbit interaction
results in different dissociation pathways.
Due to the presence of many closely located molecular states
in the $^3$D$\;-\;$$^1$D energy range
that couple to the $0^+$ and 1 symmetries, the effect of the spin-orbit
coupling is very pronounced.
Indeed, comparison of Fig. \ref{fig1} and Fig. \ref{fig5} shows that
the behavior
of some relativistic curves is drastically modified compared to
the nonrelativistic case.

In Table \ref{tab2} we report the spectroscopic constants of the
relativistic states and compare them with the available {\em ab initio}
data of Ref. \cite{Knecht:10a}. An inspection of this Table shows that the
agreement between the present results and the data of Ref.  \cite{Knecht:10a}
is excellent for the dissociations Ba($^1$S$_0$)+Rb$^+$($^1$S$_0$) and
Ba$^+$($^2$S$_{1/2}$)+Rb($^2$S$_{1/2}$). Indeed, our result for the well depth
of the ground state overestimates the data of Ref. \cite{Knecht:10a} by only 1.6\%.
For the first excited states of $0^+$ and 1 symmetry, our results underestimate
the values of Knecht et al.  \cite{Knecht:10a} by 3.1\% and 3.7\%, respectively.
For all states mentioned above, the positions of the minima in the two calculations
agree within 0.1 bohr or better. It is gratifying to observe such an excellent
agreement between two different sets of {\em ab initio} calculations performed with
different methods, CCSD(T) and RCCSD(T) in the present work vs. MRCI in Ref.
\cite{Knecht:10a}. Such a good agreement was expected from the analysis of the
nonrelativistic results, since the triple excitations are relatively unimportant
for these states. It is also worth noting that the pseudopotentials and
basis sets used in our calculations \cite{Stoll:06,Stoll:05} do a very good
job, as compared to the fully relativistic Dirac-Coulomb calculations \cite{Knecht:10a}.
The comparison for higher excited states is less favorable. For the (3)1 state
the agreement of $D_e$ is within 4\%, and $R_e$ is shifted by 0.1 bohr. This
good agreement is again not fortuitous, since the (3)1 relativistic state is
dominated by the nonrelativistic (1)$^3\Delta$ component, and the latter is
a high-spin state not very sensitive to triple excitations in the wave function.
For other states the differences in the well depths are of the order of 8\% to 13\%,
and mostly reflect the lack of triple excitations in the calculations of Ref.
\cite{Knecht:10a}. We would like to stress, however, that the overall pictures
of the relativistic states in the present paper and in Ref. \cite{Knecht:10a}
agree quite well.

\begin{figure}
\begin{center}
\begin{minipage}{\textwidth}
\epsfxsize=13cm \epsffile{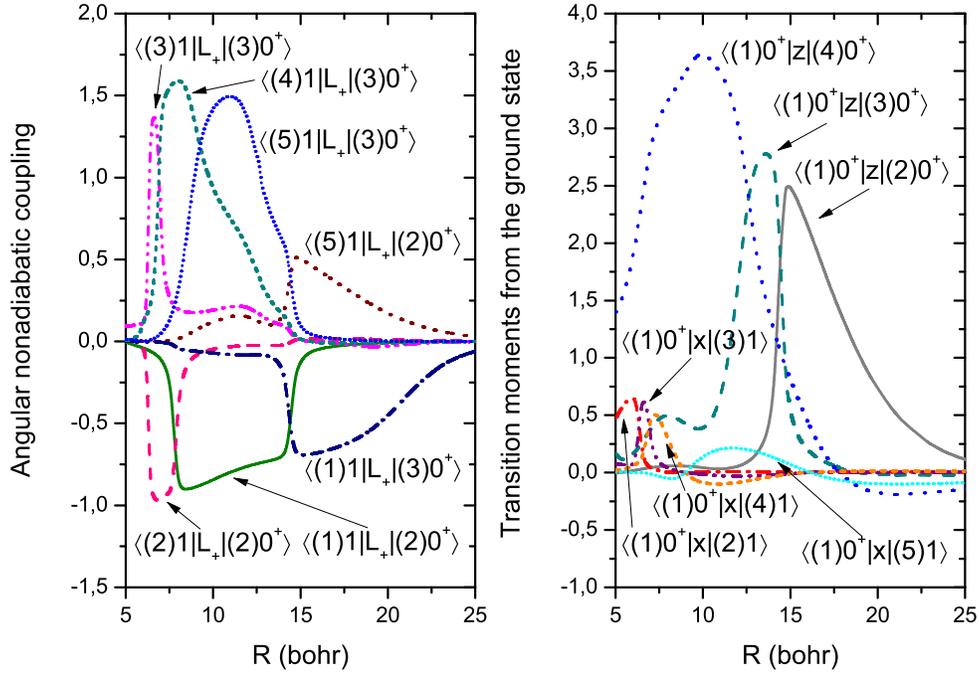}
\end{minipage}
\end{center}
\vskip 1cm
\caption{Nonadiabatic coupling matrix elements of the relativistic electronic states and
relativistic electric transition dipole moments
of the (BaRb)$^+$ molecular ion. The left and right pannels correspond to the
nonadiabatic couplings and transition moments, respectively.}
\label{fig6}
\end{figure}

As in the nonrelativistic case, in the relativistic picture states of the same
symmetry do not cross, while states of different symmetries can cross.
Given the complicated pattern of the molecular states, cf. Fig. \ref{fig5},
the knowledge of the nonadiabatic couplings is essential for the multichannel
dynamics. The nonadiabatic couplings between the relativistic states as
functions of the internuclear distance $R$ are presented on the
left pannel of Fig. \ref{fig6}.
An inspection of this figure shows that some couplings are rather localized
in space with sharp maxima or minima, and some other show broad structure.
All this can be rationalized  by looking at the predominant singlet or
triplet character of the states involved. Since all these structures
can be explained in such a way, we take
the coupling $\langle (5)1|L_+|(3)0^+\rangle$ as
an example. The (5)1 and (3)$0^+$ show crossing around $R\approx 12$ bohr, and in this region the nonadiabatic coupling has a broad maximum.
At these distances both states are primarily singlets with only a small
admixture of some triplet states. At $R\approx 15$ bohr the (3)0$^+$ state
shows an avoided crossing with the (2)0$^+$ state. In the nonrelativistic
picture the (2)0$^+$ state is mostly dominated by the (1)$^3\Pi$ state,
while the (3)0$^+$ state by the (2)$^1\Sigma$ state. Thus, at distances
larger than the avoided crossing
the matrix element of $L_+$ between the (1)$^3\Pi$ and
(1)$^1\Pi$ states is zero.

In the relativistic case many transitions that were forbidden at the
nonrelativistic level due to the different multiplicities
of the states involved become allowed
due to admixtures of singlets to triplet and vice versa,
cf. the right pannel
of Fig. \ref{fig6}. These additional transition moments are very small,
showing that the relativistic states obtained by admixture of the singlet
states to triplets are almost pure triplet states. The transition moment
$\langle (1)0^+|z|(4)0^+\rangle$ resembles very much
$\langle (1)^1\Sigma|z|(3)^1\Sigma\rangle$, with little differences only
at small internuclear distances. More interesting are the transition
moments $\langle (1)0^+|z|(2)0^+\rangle$ and $\langle (1)0^+|z|(3)0^+\rangle$.
If one would take the sum of these two curves, the resulting
curve would strongly ressemble
the graph for $\langle (1)^1\Sigma|z|(2)^1\Sigma\rangle$ transition moment,
cf. Fig. \ref{fig3}.
The explanation of this fact is very simple. At distances smaller than
$R\approx 15$ bohr the (3)$0^+$ is predominantly a singlet state, while
(4)$0^+$ has mostly triplet character. The situation is opposite at $R$
larger than 15 bohr. This explains why we get two curves with a sharp
decay to zero around 15 bohr.

\subsection{Long-range behavior of the nonrelativistic and relativistic potentials}
\label{sub6}
\begin{figure}
\begin{center}
\begin{minipage}{\textwidth}
\epsfxsize=13cm \epsffile{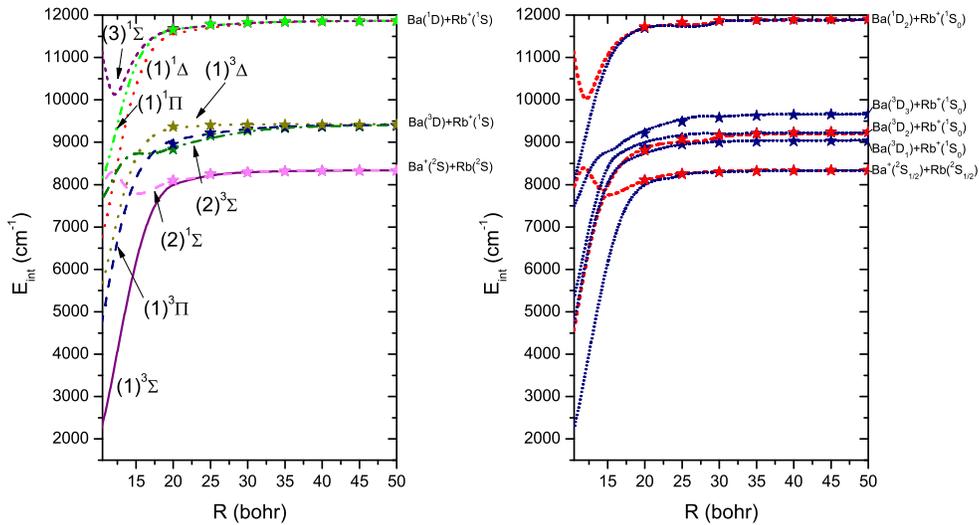}
\end{minipage}
\end{center}
\vskip 1cm
\caption{Long range nonrelativistic (left pannel) and relativistic (right pannel) potential energy curves
of the (BaRb)$^+$ molecular ion. Stars are based on long range coefficients like in Eq. \ref{lr2}.}
\label{fig6a}
\end{figure}

When describing cold collisions between atoms it is crucial to ensure the proper
long-range asymptotics of the interaction potential. The long-range coefficients
describing the asymptotics of the nonrelativistic potentials are reported in
Table \ref{tab3}. An inspection of this Table shows that dispersion contribution
$C_6^{\rm disp}$ is modest, but not negligible, of the order of 5\% to 20\%.
The induction and dispersion coefficients are always positive, so they describe
the attractive parts of the potentials. The electrostatic coefficients differ
in sign depending on the state considered and are responsible for the appearance
of barriers and long-range minima on the potential energy curves. This is
illustrated on the left pannel of Fig. \ref{fig6a}, where it is shown how the
long-range asymptotics nicely fits the {\em ab initio} points. It is worth noting
that the quadrupole moments are extremely sensitive to the electronic correlation. For
instance, the value of $C_3^{\rm elst}$ for the (3)$^1\Sigma$ state is 1.16
at the LRCC3 level and 2.05 with the LRCCSD method. This means that the
triple excitations diminish $C_3^{\rm elst}$ by as much as 44\%. Surprisingly
enough, such an effect is not observed for the hexadecapole moment. Here,
the $C_5^{\rm elst}$ at the LRCC3 and LRCCSD differ by only 6\%.

The long-range coefficients describing the large $R$ asymptotics of the
relativistic states are presented in Table \ref{tab4}. As discussed in \ref{sub5}
in the atomic limit there is no spin-orbit coupling between the
ground X$^1\Sigma$, (1)$^3\Sigma$, and (2)$^1\Sigma$ states with other
states, so the long-range coefficients of the (1)$0^+$, (1)1, and (2)$0^+$ states
remain unchanged in this approximation. An inspection of Table \ref{tab4} shows
the SO coupling of the $^3$D and $^1$D states modestly affects the long-range
coefficients in the Ba($^1$D$_2$)+Rb$^+$($^1$S$_0$) dissociation limit. The SO
coupling in the $^3$D multiplet introduces larger changes. For instance, the
$C_5^{\rm elst}$ coefficient for the (2)1 state is zero. This is not fortuitous,
but only reflects the fact that the hexadecapole moment of an atom in a $^3$D$_1$
state is identically zero. Again, the signs of the electrostatic coefficients
are responsible for the barriers and long-range minima, cf. the right pannel
of Fig. \ref{fig6a}.

\subsection{Elastic cross sections, spin-exchange, and radiative charge transfer}
\label{sub7}
\begin{figure}
\begin{center}
\begin{minipage}{\textwidth}
\epsfxsize=13cm \epsffile{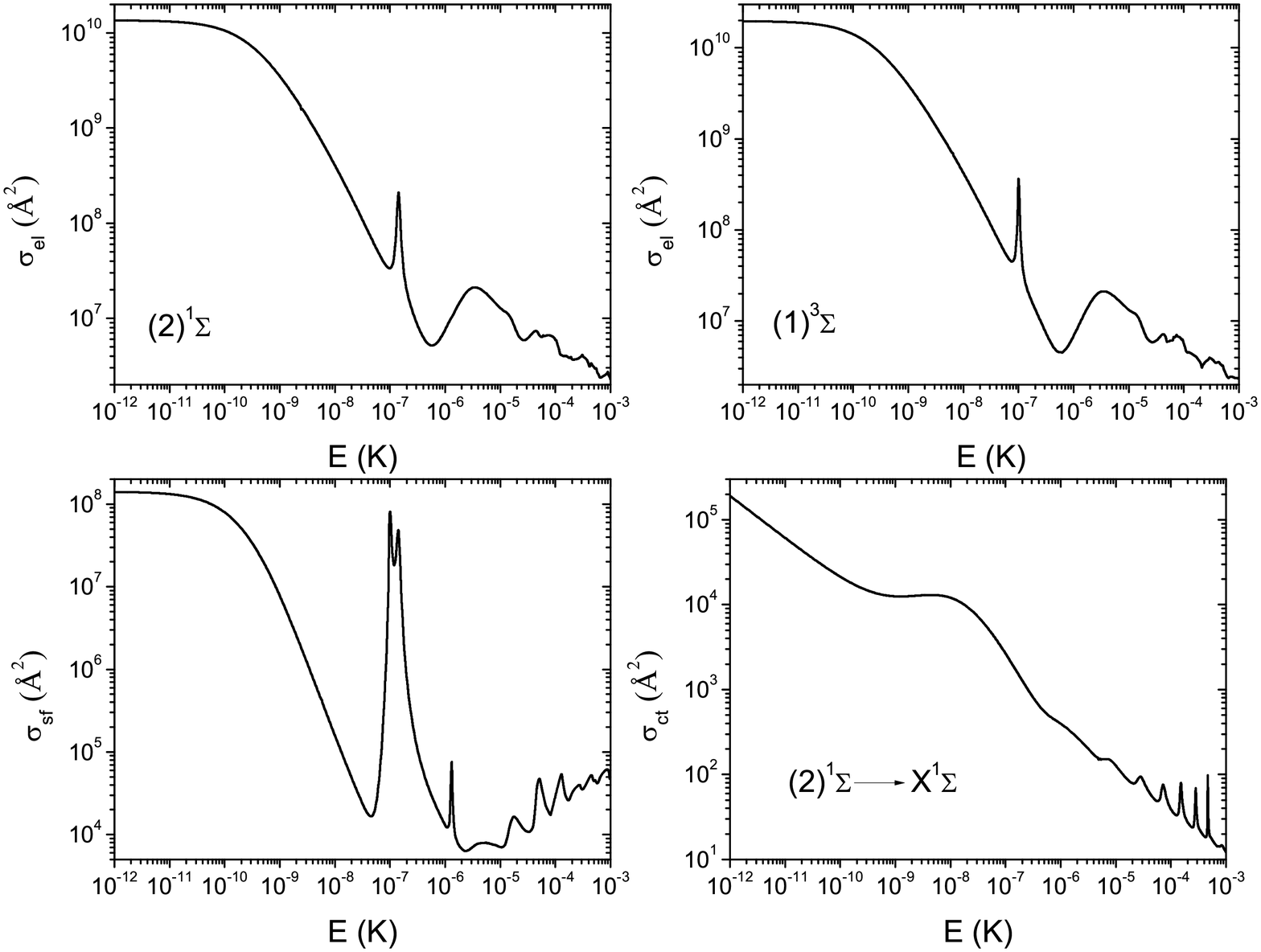}
\end{minipage}
\end{center}
\vskip 5cm
\caption{Elastic, spin-flip, and charge transfer cross sections for collisions of
${}^{138}$Ba$^+$($^2$S) and ${}^{87}$Rb($^2$S) as functions of the collision energy from nonrelativistic
potentials.}
\label{fig7}
\end{figure}

Thus far we have discussed the results of the electronic structure calculations.
Now, we turn to the problem of sympathetic cooling of cold barium ions, and present
the results for the elastic, spin-flip, and charge transfer cross sections
for both the nonrelativistic and relativistic potentials.
In Fig. \ref{fig7} we report the elastic cross sections in the singlet and
triplet manifolds, the spin-flip cross section, and the charge transfer
cross section from the (2)$^1\Sigma$ to ground X$^1\Sigma$ state, all
calculated from the nonrelativistic potentials. An inspection of this
figure shows that in the ultracold regime the elastic cross sections
behave according to the Wigner's threshold law, and the value of the cross
section extrapolated to zero energy agrees very well with the one determined
from the $s$ wave scattering length. At energies around 100 nK the energy
dependence of the cross sections starts to exhibit some structures related to
shape resonances appearing due to the contributions of higher partial waves
in the expansions (\ref{el}) and to glory interference effects.
Note that in the range of energies up to 1 mK the curves representing
$\sigma_{\rm el}^{\rm s}(E)$ and $\sigma_{\rm el}^{\rm t}(E)$ are
hardly distinguishable, despite the fact that the potential energy
curves for the (2)$^1\Sigma$ and (1)$^3\Sigma$ states are quite
different. This behavior is purely fortuitous, and is due to the fact
that these two states have the same asymptotics and very close
scattering lengths, $a_{\rm s}$ and $a_{\rm t}$, equal to $-3.53\times 10^5$ and
$-4.26 \times 10^5$ \AA, respectively. The superelastic spin-flip cross section shows
a qualitatively similar behavior. Overall, all the elastic cross sections
are very large, from around 10$^{10}$ \AA$^2$ at ultralow temperatures
to $\approx$10$^6$ \AA$^2$ in the mK region.

In the ultacold regime the charge transfer cross section, which is
an inelastic state changing cross section, decays as $E^{-1/2}$, in
accordance with the Wigner's threshold law. At nanoKelvin temperatures
this cross section, of the order of 10$^4$ \AA$^2$ is five orders of
magnitude smaller than the elastic cross sections. When we go up
to the milliKelvin temperatures this ratio is even slightly more
favorable, about five orders of magnitude of difference. Thus we can
conclude that at the nonrelativistic level and with the single
channel description of the collisional dynamics, cooling of the
barium ion by collisions with ultracold rubidium atoms should be
very efficient.

\begin{figure}
\begin{center}
\begin{minipage}{\textwidth}
\epsfxsize=13cm \epsffile{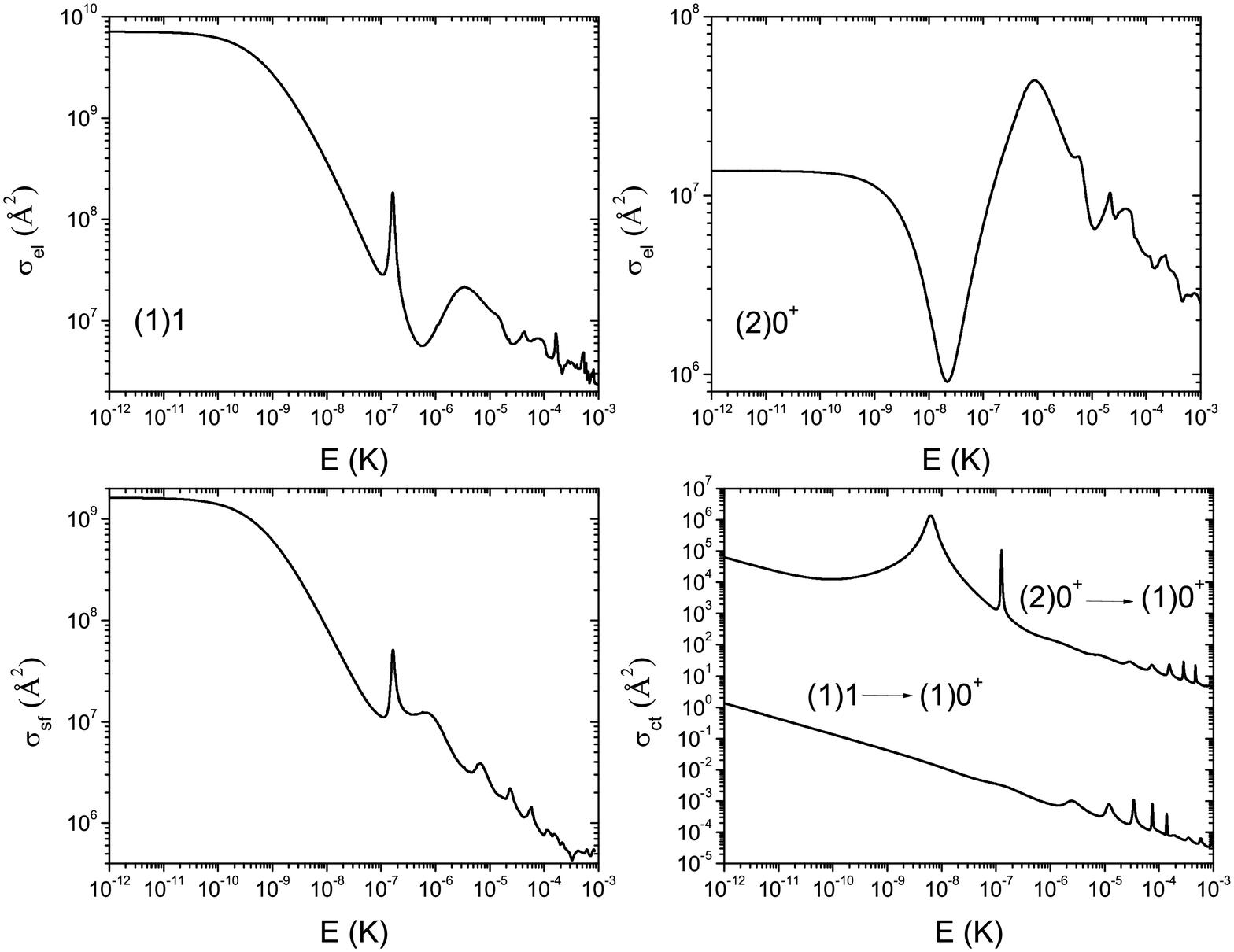}
\end{minipage}
\end{center}
\vskip 5cm
\caption{Elastic, spin-flip, and charge transfer cross sections for collisions of
${}^{138}$Ba$^+$($^2$S) and ${}^{87}$Rb($^2$S) as functions of the collision energy from relativistic
potentials.}
\label{fig8}
\end{figure}

The results of dynamical calculations on the relativistic (2)$0^+$
and (1)1 potentials are presented on Fig. \ref{fig8}. The elastic
cross section from the (1)1 state is almost indentical to the
cross section obtained with the nonrelativistic (1)$^3\Sigma$
potential. This is not surprising since the spin-orbit coupling
has a very small effect on the potential, slightly moving the
repulsive wall, cf. Figs. \ref{fig1} and \ref{fig5}. The
relativistic scattering length is lower compared to the
nonrelativistic (1)$^3\Sigma$ value, but most of the features,
resonance structure and glory interference effects, are almost
the same. By contrast, the energy dependence of the elastic
cross section in the (2)$0^+$ potential is very different from
that presented in Fig. \ref{fig7} for the nonrelativistic
potential. We note that the Wigner's limit is very different,
and the resonant structure, glory ondulation, and all details
at higher energies changed drastically. For instance, the
scattering lengths for the (2)$0^+$ and (1)1 states are of
$-1.58\times 10^3$ and $-2.56\times 10^5$ \AA, to be compared with the values of
$a_{\rm s}$ and $a_{\rm t}$, $-4.26\times 10^5$ and $-3.53\times 10^5$ \AA, quoted above.
All these difference are not surprising,
however, since the spin-orbit mixing has a profound effect in this
case. In fact, up to $R\approx 15$ bohr, the point of the
avoided crossing between the potential energy curves of the (2)$0^+$
and (3)$0^+$ states, the (2)$0^+$ potential has predominantly
the character of the (1)$^3\Pi$ state. Only after the avoided
crossing and mostly in the long range it becomes an almost
pure (2)$^1\Sigma$ state. Also the spin-flip cross section
computed from the relativistic potentials  is quite different
from the nonrelativistic one. It shows a sharp resonant
structure around in the $\mu$K region.

The stricking difference between the nonrelativistic and relativistic
pictures is the fact that the charge transfer process is now allowed from
both the (2)$0^+$ and (1)1 states. The corresponding cross sections
as functions of the energy are also shown on Fig. \ref{fig7}.
Again, we observe that the Wigner's threshold law with the $E^{-1/2}$
decay of the cross section in the ultracold regime is preserved.
An inspection of this figure shows that the charge transfer process
from the (1)1 state to the ground state will be very slow, unimportant
at temperatures interesting from the experimental point of view.
This is not a surprise, since the transition dipole moment from
the (1)1 state to the ground state is very small, cf. Fig. \ref{fig6}.
The charge transfer cross section from the (2)$0^+$ state is
significantly larger, but again is several orders of magnitude
smaller than the elastic one over all the range of temperatures
considered in our work. In particular, in the milliKelvin regime,
the inelastic events are six orders of magnitude less probable than
the elastic. Thus, the relativistic calculations confirm the
conclusions from the nonrelativistic case that the sympathetic cooling
of the barium ion by collisions with ultracold rubidium atoms will
be very efficient.

We would like to conclude this section by saying that the present
single channel calculations strongly suggest that the sympathetic
cooling of the barium atom by collisions with ultracold rubidium
atoms should be very efficient in the temperature range up to
milliKelvin. The analysis presented here neglects the effect of the hyperfine splittings which are important at low collision energies. Thus, our single channel analysis performed in terms of singlet and triplet properties is only approximate at ultracold temperatures, and the actual collision rates will depend on the hyperfine spin states of the atom and ion. Moreover, other inelastic processes due to the
presence of other channels could enter the game, and final
conclusions can be presented only when full multichannel
calculations are performed. Work in direction is in progress
\cite{Simoni:10}.

\section{Summary and conclusions}
\label{sec4}
In this paper we have reported theoretical prospects for the symphatetic cooling of the barium ion by collisions with the
ultracold buffer gas of rubidium atoms. Potential energy curves for the ground X$^1\Sigma$ state of the (BaRb)$^+$ molecular ion
corresponding to the Rb$^+$($^1$S)+Ba($^1$S) dissociation and for the excited states, $(1)^3\Sigma$ and $(2)^1\Sigma$,
corresponding to the
Rb($^2$S)+Ba$^+$($^2$S) dissociation were computed by means of size-consistent coupled cluster methods with
single, double, and triple excitations in the wave function. The inclusion of the triple excitations in the wave function
was shown to be essential for the open-shell $(2)^1\Sigma$ state. The (BaRb)$^+$ molecular ion shows a lot of
low-lying molecular states, and the corresponding potential energy curves were computed as well.
Using the molecular spin-orbit coupling matrix elements relativistic potential energy curves were obtained.
The asymptotics of the nonrelativistic and relativistic potentials was fixed with long-range coefficients calculated from
$C_3$ up to and including $C_6$.
A good understanding of the dynamics in the (BaRb)$^+$ system requires the knowledge of the nonadiabatic (radial and angular)
coupling matrix elements, which were calculated as well. Finally, the electric dipole transition moments from the ground state
were computed. Based on the above {\em ab initio} electronic structure calculations, single channel dynamical
calculations of the elastic, spin-exchange, and inelastic cross sections for the collisions of the Ba$^+$($^2$S) ion with
the Rb($^2$S) in the energy range from 0 to 1 mK were performed, using both the nonrelativistic and relativistic potentials.
It was found that the elastic processes are a few orders of magnitude more favorable that the inelastic ones.
Thus, we can conclude our paper by saying that the sympathetic cooling of the barium ion by the buffer gas of ultracold rubidium
atoms should be very efficient taking into account the two-body collisional properties of Rb and Ba$^+$.

%Since the sympathetic cooling does not interfere with the internal degrees of freedom it should be possible to encode quantum information states on barium ions and constantly cool them by collisions with rubidium.

\section*{Acknowledgments}
We would like to
thank Agnieszka Witkowska for fruitful discussions, and for reading and commenting on the manuscript. This work was supported by
the Polish Ministry of Science and Higher Education
(grants 1165/ESF/2007/03 and PBZ/MNiSW/07/2006/41),
and the Foundation for Polish Science (FNP)
via Homing program (grant no. HOM/2008/10B) within the EEA Financial
Mechanism.

\newpage

\begin{table}[h]
\caption{Spectroscopic characteristics of the nonrelativistic electronic states
of the (BaRb)$^+$ molecular ion.\label{tab1}}
\vskip 3ex
\begin{tabular}{lrrrll}
\hline\hline
State                          & $R_e$ (bohr) & $D_e$ (cm$^{-1}$) & $T_e$ (cm$^{-1}$)& Reference & Dissociation \\
\hline
X$^1\Sigma$                    &  8.67 &   5136    &       0   & present              & Ba($^1$S)+Rb$^+$($^1$S) \\
                               &  8.85 &           &       0   & Ref. \cite{Knecht:10}&                         \\
\hline
(1)$^3\Sigma$                  &  9.27 &   6587    &    6893   & present              & Ba$^+$($^2$S)+Rb($^2$S) \\
                               &  9.38 &           &    6808   & Ref. \cite{Knecht:10}&                         \\
(2)$^1\Sigma$ (primary minimum) &  9.02 &    911    &   12569   & present              & Ba$^+$($^2$S)+Rb($^2$S) \\
(2)$^1\Sigma$ (secondary minimum)&15.19 &    576    &   12904   & present              & Ba$^+$($^2$S)+Rb($^2$S) \\
\hline
(2)$^3\Sigma$ (primary minimum) &  9.82 &   1874    &   12683   & present              & Ba($^3$D)+Rb$^+$($^1$S) \\
                               &  9.58 &           &   13478   & Ref. \cite{Knecht:10}&                         \\
(2)$^3\Sigma$ (secondary minimum)&16.78 &    697    &   13861   & present              & Ba($^3$D)+Rb$^+$($^1$S) \\
(1)$^3\Pi$                     &  8.19 &  6301 &   8257   & present              & Ba($^3$D)+Rb$^+$($^1$S) \\
                               &  8.17 &           &    8832   & Ref. \cite{Knecht:10}&                         \\
(1)$^3\Delta$                  &  9.11 &   4380    &   10178   & present              & Ba($^3$D)+Rb$^+$($^1$S) \\
                               &  9.19 &           &   10776   & Ref. \cite{Knecht:10}&                         \\
\hline
(3)$^1\Sigma$                  & 12.00 &   1858    &   15146   & present              & Ba($^1$D)+Rb$^+$($^1$S) \\
(1)$^1\Pi$                     &  9.04 &   4403    &   12601   & present              & Ba($^1$D)+Rb$^+$($^1$S) \\
                               &  8.44 &           &   15906   & Ref. \cite{Knecht:10}&                         \\
(1)$^1\Delta$                  &  9.08 &   5769    &   11236   & present              & Ba($^1$D)+Rb$^+$($^1$S) \\
                               &  9.19 &           &   11276   & Ref. \cite{Knecht:10}&                         \\
\hline\hline
\end{tabular}
\end{table}

\newpage

\begin{table}[h]
\caption{Spectroscopic characteristics of the relativistic electronic states
of the (BaRb)$^+$ molecular ion.\label{tab2}}
\vskip 3ex
\begin{tabular}{lrrrll}
\hline\hline
State                          & $R_e$ (bohr) & $D_e$ (cm$^{-1}$) & $T_e$ (cm$^{-1}$)& Reference & Dissociation \\
\hline
(1)$0^+$                       &  8.67 &   5136    &       0   & present              & Ba($^1$S$_0$)+Rb$^+$($^1$S$_0$) \\
                               &  8.72 &   5055    &       0   & Ref. \cite{Knecht:10a}&                        \\
\hline
(1)1                           &  9.25 &   6609    &    6872   & present              & Ba$^+$($^2$S$_{1/2}$)+Rb($^2$S$_{1/2}$) \\
                               &  9.22 &   6871    &    6638   & Ref. \cite{Knecht:10a}&                         \\
(2)$0^+$                       &  8.17 &   5403    &    8077   & present              & Ba$^+$($^2$S$_{1/2}$)+Rb($^2$S$_{1/2}$) \\
                               &  8.28 &   5899    &    7775   & Ref. \cite{Knecht:10a}&                         \\
\hline
(2)1                           &  8.27 &   5878    &    8293   & present              & Ba($^3$D$_1$)+Rb$^+$($^1$S$_0$) \\
                               &  8.28 &   5742    &    7932   & Ref. \cite{Knecht:10a}&                         \\
(3)1                           &  9.10 &   4497    &    9893   & present              & Ba($^3$D$_2$)+Rb$^+$($^1$S$_0$) \\
                               &  9.22 &   4302    &    9556   & Ref. \cite{Knecht:10a}&                         \\
(4)1                           &  9.41 &   2353    &   12464   & present              & Ba($^3$D$_3$)+Rb$^+$($^1$S$_0$) \\
                               &  9.22 &   2258    &   11963   & Ref. \cite{Knecht:10a}&                         \\
(3)$0^+$                       &  9.01 &   1801    &   12589   & present              & Ba($^3$D$_2$)+Rb$^+$($^1$S$_0$) \\
                               &  9.03 &           &   12005   & Ref. \cite{Knecht:10a}&                         \\
\hline
(5)1                           &  9.45 &   4108    &   12936   & present              & Ba($^1$D$_2$)+Rb$^+$($^1$S$_0$) \\
                               &  9.72 &   3556    &   12687   & Ref. \cite{Knecht:10a}&                         \\
(4)$0^+$                       & 12.01 &   1890    &   15153   & present              & Ba($^1$D$_2$)+Rb$^+$($^1$S$_0$) \\
\hline\hline
\end{tabular}
\end{table}

\newpage

\begin{table}[h]
\caption{Long-range coefficients (in atomic units) for the nonrelativistic electronic states
of the (BaRb)$^+$ molecular ion. $C_6$ is the sum $C_6^{\rm ind}+C_6^{\rm disp}$.\label{tab3}}
\vskip 3ex
\begin{tabular}{lrrrrrrl}
\hline\hline
State                          & $C_3^{\rm elst}$  & $C_4^{\rm ind}$ & $C_5^{\rm elst}$&  $C_6^{\rm ind}$ &
$C_6^{\rm disp}$ & $C_6$ & Dissociation \\
\hline
X$^1\Sigma$                    &       &    136.8  &         & 4450 & 368 & 4818   & Ba($^1$S)+Rb$^+$($^1$S) \\
\hline
(1)$^3\Sigma$                  &       &    159.3  &         & 3260 &2510 & 5770   & Ba$^+$($^2$S)+Rb($^2$S) \\
(2)$^1\Sigma$                  &       &    159.3  &         & 3260 &2510 & 5770   & Ba$^+$($^2$S)+Rb($^2$S) \\
\hline
(2)$^3\Sigma$                  &--4.55 &   324.1   &   106.3 & 1873 & 392 & 2265    & Ba($^3$D)+Rb$^+$($^1$S) \\
(1)$^3\Pi$                     &--2.27 &   272.0   &  --71.0 & 2718 & 372 & 3090    & Ba($^3$D)+Rb$^+$($^1$S) \\
(1)$^3\Delta$                  &  4.55 &   115.1   &    18.1 & 3060 & 322 & 3382    & Ba($^3$D)+Rb$^+$($^1$S) \\
\hline
(3)$^1\Sigma$                  &--1.16 &   108.1   &  154.3  & 2844 & 661 & 3505    & Ba($^1$D)+Rb$^+$($^1$S) \\
(1)$^1\Pi$                     &--0.58 &   127.0   &--102.9  & 2878 & 244 & 3122    & Ba($^1$D)+Rb$^+$($^1$S) \\
(1)$^1\Delta$                  &  1.16 &   180.9   &   25.7  & 3802 & 273 & 4075    & Ba($^1$D)+Rb$^+$($^1$S) \\
\hline\hline
\end{tabular}
\end{table}

\newpage

\begin{table}[h]
\caption{Long-range coefficients (in atomic units) for the relativistic electronic states
of the (BaRb)$^+$ molecular ion.\label{tab4}}
\vskip 3ex
\begin{tabular}{lrrrrl}
\hline\hline
State                          & $C_3^{\rm elst}$  & $C_4^{\rm ind}$ & $C_5^{\rm elst}$&  $C_6$ & Dissociation \\
\hline
(1)$0^+$                       &       &    136.8  &         & 4818 & Ba($^1$S$_0$)+Rb$^+$($^1$S$_0$) \\
\hline
(1)1                           &       &    159.3  &         & 5770 & Ba$^+$($^2$S$_{1/2}$)+Rb($^2$S$_{1/2}$)\\
(2)$0^+$                       &       &    159.3  &         & 5770 & Ba$^+$($^2$S$_{1/2}$)+Rb($^2$S$_{1/2}$)\\
\hline
(2)1                           &  1.60 &   183.0   &     0.0 & 3183 & Ba($^3$D$_1$)+Rb$^+$($^1$S$_0$) \\
(3)1                           &--1.13 &   244.0   &    46.0 & 2780 & Ba($^3$D$_2$)+Rb$^+$($^1$S$_0$) \\
(4)1                           &--2.73 &   282.0   &     6.0 & 3382 & Ba($^3$D$_3$)+Rb$^+$($^1$S$_0$) \\
(3)$0^+$                       &--2.25 &   270.0   &  --66.7 & 3096 & Ba($^3$D$_2$)+Rb$^+$($^1$S$_0$) \\
\hline
(5)1                           &--0.59 &   129.0   &--101.0  & 3117 & Ba($^1$D$_2$)+Rb$^+$($^1$S$_0$) \\
(4)$0^+$                       &--1.18 &   110.0   &  151.0  & 3493 & Ba($^1$D$_2$)+Rb$^+$($^1$S$_0$) \\
\hline\hline
\end{tabular}
\end{table}

\newpage
\newpage
\newpage
\newpage
\newpage
\newpage
\newpage
\newpage
\newpage
\end{document}